\documentclass[natbib,graybox]{svmult}
\usepackage{mathptmx}       % selects Times Roman as basic font
\usepackage{helvet}         % selects Helvetica as sans-serif font
\usepackage{courier}        % selects Courier as typewriter font
\usepackage{type1cm}        % activate if the above 3 fonts are
\usepackage{makeidx}         % allows index generation
\usepackage{graphicx}        % standard LaTeX graphics tool
\usepackage{multicol}        % used for the two-column index
\usepackage[bottom]{footmisc}% places footnotes at page bottom
\usepackage{hyperref}
\hypersetup{
    colorlinks,%
    citecolor=blue,%
    linkcolor=blue,%
    urlcolor=blue
}
\usepackage{amssymb,amsmath}
\newcommand{\be}{\begin{equation}}
\newcommand{\unit}[1]{\nobreak{\mathrm{\;#1}}} % For units of measure within math mode
\newcommand{\ee}{\end{equation}}
\newcommand{\beq}{\begin{eqnarray}}

\def\comp{\,c/\omega_{\rm p}}
\newcommand{\bmath}[1]{\mbox{\boldmath{$#1$}}}
\def\ompt{\omega_{\rm p}t}
\newcommand{\eeq}{\end{eqnarray}}
\newcommand\subsun[1]{{$_{\normalsize\odot}$}}
\newcommand\sect[1]{Sect.~\ref{#1}}
\newcommand\fig[1]{Fig.~\ref{fig:#1}}

\newcommand{\rj}{\,r_{\rm j}}
\newcommand{\ie}{\emph{i.e.} }

\newcommand{\sigmain}{\sigma_{\rm in}}
\newcommand{\rhot}{r_{\rm L,hot}}
\newcommand{\gammamax}{\gamma_{\rm max}}

\makeindex             % used for the subject index
\begin{document}

%\frontmatter%%%%%%%%%%%%%%%%%%%%%%%%%%%%%%%%%%%%%%%%%%%%%%%%%%%%%%

%\include{dedic}
%\include{foreword}

%\include{preface/preface} TO ADD AT THE END (including soc, etc.)

%\include{acknow} TO ADD AT THE END (including soc, etc.)

%\tableofcontents

%\include{cblist}

\title{Particle Acceleration in Pulsar Wind Nebulae: PIC modelling}%\thanks{Both authors contributed equally to this chapter.}
\author{Lorenzo Sironi and Beno\^it Cerutti}
\institute{Both authors contributed equally to this chapter. \at \at Lorenzo Sironi \at Columbia University, Pupin Hall, 550 West 120th Street,
New York, NY 10027, USA. \email{lsironi@astro.columbia.edu}
\and Beno\^it Cerutti \at Univ. Grenoble Alpes, CNRS, IPAG, F-38000 Grenoble, France. \email{benoit.cerutti@univ-grenoble-alpes.fr}}
\maketitle

\abstract*{We discuss the role of PIC simulations in unveiling the origin of the emitting particles in PWNe. After describing the basics of the PIC technique, we summarize its implications for the quiescent and the flaring emission of the Crab Nebula, as a prototype of PWNe. A consensus seems to be emerging that, in addition to the standard scenario of particle acceleration via the Fermi process at the termination shock of the pulsar wind, magnetic reconnection in the wind, at the termination shock and in the nebula plays a major role in powering the multi-wavelength emission signatures of PWNe.}

\abstract{We discuss the role of PIC simulations in unveiling the origin of the emitting particles in PWNe. After describing the basics of the PIC technique, we summarize its implications for the quiescent and the flaring emission of the Crab Nebula, as a prototype of PWNe. A consensus seems to be emerging that, in addition to the standard scenario of particle acceleration via the Fermi process at the termination shock of the pulsar wind, magnetic reconnection in the wind, at the termination shock and in the Nebula plays a major role in powering the multi-wavelength signatures of PWNe.}

\section{Introduction}\label{sect_intro}
In recent years, multi-dimensional magnetohydrodynamic (MHD) models of Pulsar Wind Nebulae (PWNe) have been able to reproduce the nebular morphology down to intricate details (see contributions by A. Mignone, L. del Zanna and O. Porth in this volume). In order to compare the results of MHD simulations to the multi-wavelength observations of PWNe (most notably, of the prototypical Crab Nebula), it is usually assumed that the termination shock of the pulsar wind --- where the ram pressure of the ultra-relativistic wind emanating from the pulsar balances the thermal pressure of the surrounding nebula --- is an efficient site of particle acceleration. This assumption cannot be tested directly within the MHD framework (which bears no information on the properties of the accelerated particles), but it requires fully-kinetic particle-in-cell (PIC) simulations. By capturing the interplay of charged particles and electromagnetic fields from first principles, PIC simulations allow to identify potential locations of particle acceleration in PWNe. On the other hand, due to computational constraints, PIC simulations are typically confined to a local description of the system, on scales much smaller than the nebular size. It is only by integrating the PIC results with a global MHD model of the nebula that we can properly reproduce the multi-wavelength signatures of PWNe.

In this chapter, we summarize the role of PIC simulations in understanding the origin of high-energy particles in PWNe. In Sect.~\ref{sect_methods} we describe the basics of the PIC method. In Sect.~\ref{sect_applications} we summarize the implications of PIC results on the quiescent and flaring emission from PWNe. We conclude in Sect.~\ref{sect_conc}.

\section{The particle-in-cell technique}\label{sect_methods}

This section is intended to provide a brief overview of the most common methods and algorithms used in explicit PIC codes. A detailed presentation of this technique can be found in \citet{1988csup.book.....H, 1991ppcs.book.....B}.

\subsection{Collisionless plasmas}

A necessary condition for non-thermal particle acceleration is the absence of Coulomb collision in the plasma of interest. This is the case for most high-energy astrophysical systems, and in particular pulsar wind nebulae, where plasmas are very diluted. Roughly speaking, a plasma can be considered as ``collisionless'' if the frequency of Coulomb collision ($\nu$) is much smaller than the plasma frequency, $\omega_{\rm pe}\gg \nu$. This condition implies that the number of particles per Debye sphere must be large, i.e., $N_{\rm D}\gg 1$. The dynamics of individual particle is driven by collection plasma phenomena rather than binary collisions at the sub-Debye length and plasma frequency scales, or simply referred below as the ``kinetic'' scale. As we will see in Section~\ref{sect_applications}, these microscopic scales are involved in the particle acceleration processes and, thus, they must be well-resolved by simulations in contrast to the magnetohydrodynamic approach. To obtain meaningful astrophysical results, particle-based simulations must also capture large scale features, i.e., system size and long integration time.

The evolution of a collisionless plasma is governed by the Vlasov equation
\begin{equation}
\frac{\partial f}{\partial t}+\frac{\mathbf{p}}{\gamma m}\cdot\frac{\partial f}{\partial \mathbf{r}}+q\left(\mathbf{E}+\frac{\mathbf{v}\times\mathbf{B}}{c}\right)\cdot\frac{\partial f}{\partial \mathbf{p}}=0,
\end{equation}
where $f\equiv d{\rm N}/d\mathbf{r}d\mathbf{p}$ is the particle distribution function defined in 6D phase space $\left(\mathbf{r},\mathbf{p}\right)$ and 1D in time, with $\mathbf{r}$ is the position and $\mathbf{p}=\gamma m\mathbf{v}$ is the momentum, and $q$ is the electric charge. Along with Maxwell's equations for the fields ($\mathbf{E}$ and ${\mathbf{B}}$), this is the full set of equations to model a collisionless plasma from first principles.

\subsection{The particle approach}

Analytical solutions to the Vlasov equation are known for a few idealized situations only. In most cases, it must be solved numerically. They are at least two ways to solve this equation. In the first approach, phase space is treated as a continuous fluid and Vlasov equation is solved directly using semi-Lagrangian or Eulerian methods \citep{1976JCoPh..22..330C, 2006JCoPh.213..862E}. This method has the advantage to be insensitive to particle noise, and hence can capture well weak plasma phenomena and broad particle distribution functions. In theory this is the most appropriate approach to follow, but in practice the usage of Vlasov codes is currently limited due to prohibitive numerical costs for multidimensional problems (6D). The second approach is the PIC method which is the main focus of this chapter.

In PIC, Vlasov equation is solved indirectly by integrating discrete particle trajectories. This approach is equivalent to the direct method, and an easy way to see this is to rewrite Vlasov equation as a usual advection equation: $\partial f/\partial t+\boldsymbol{\nabla}_{\mathbf{r,p}}\cdot \left(f\mathbf{U}\right) =\mathbf{0}$, where $\boldsymbol{\nabla}_{\mathbf{r,p}}=\left(\partial/\partial\mathbf{r},\partial/\partial\mathbf{p}\right)$ and $\mathbf{U}=\left(\mathbf{p}/\gamma m, q\left(\mathbf{E}+\mathbf{v}\times\mathbf{B}/c\right)\right)$. Thus, using the methods of characteristics, this first-order partial differential equation can be rewritten as a sets of ordinary differential equations (Newton's law) along characteristic curves which corresponds here to particle trajectories. For point-like particles, the particle distribution function is then approximated as
\begin{equation}
f\left(\mathbf{r},\mathbf{p},t\right)\approx\sum_{k=1}^{N}w_{k}\delta\left(\mathbf{r}-\mathbf{r}_{k}(t)\right) \delta\left(\mathbf{p}-\mathbf{p}_{k}(t)\right),
\end{equation}
where $\delta$ is the Dirac delta function and $w_k$ is the particle weight. The number of particles must be very high for a good sampling of phase space and to be close to the exact solution of Vlasov equation. In practice, however, this number will be limited by computing resources and is always much smaller than the number of particles contained in real plasmas. To overcome this difficulty, a PIC particle represents a large number (given by the weight $w_k$) of physical particles that would follow the same trajectory in phase space (with the same $q/m$ ratio). For this reason, the simulation particles are usually called ``macroparticles''.

Even though the plasma is collisionless, particles feels each other via long-range interactions. Summing over all particle-particle binary interactions, i.e. $N\left(N-1\right)/2\approx N^2$, is numerically expansive and hard to implement. Instead, in PIC, particles do not feel each other directly but via the electromagnetic fields known on the grid which result from the evolution of the plasma. In this case, the number of operations scales as the number of particles $N$ instead of $N^2$.

The PIC method has become increasingly popular in high-energy astrophysics to model non-thermal particle acceleration phenomena. PIC codes are much cheaper in comparison to Vlasov codes, and they are also conceptually simple, robust and easy to implement and parallelize efficiently to a large number of cores. This simplicity comes at the cost of significant particle noise which can lead to poor sampling of the particle distribution (e.g., steep power-law tails), difficulty in capturing subtle or weak phenomena, artificial collisions, and load-balancing issues in parallel computing.

\subsection{Main computing procedures in PIC}\label{sect_loop}

Figure~\ref{fig_loop} describes the three main operations performed per timestep $\Delta t$ of an explicit PIC code: (i) Solve Newton's equation for each particle to evolve velocities and positions (ii) Collect charge and current densities from all particles and deposit them on the grid, and (iii) Solve Maxwell's equations to update the fields on the grid. Below is a brief technical description of each step:

\begin{figure}[]
\centering
\includegraphics[width=8.5cm]{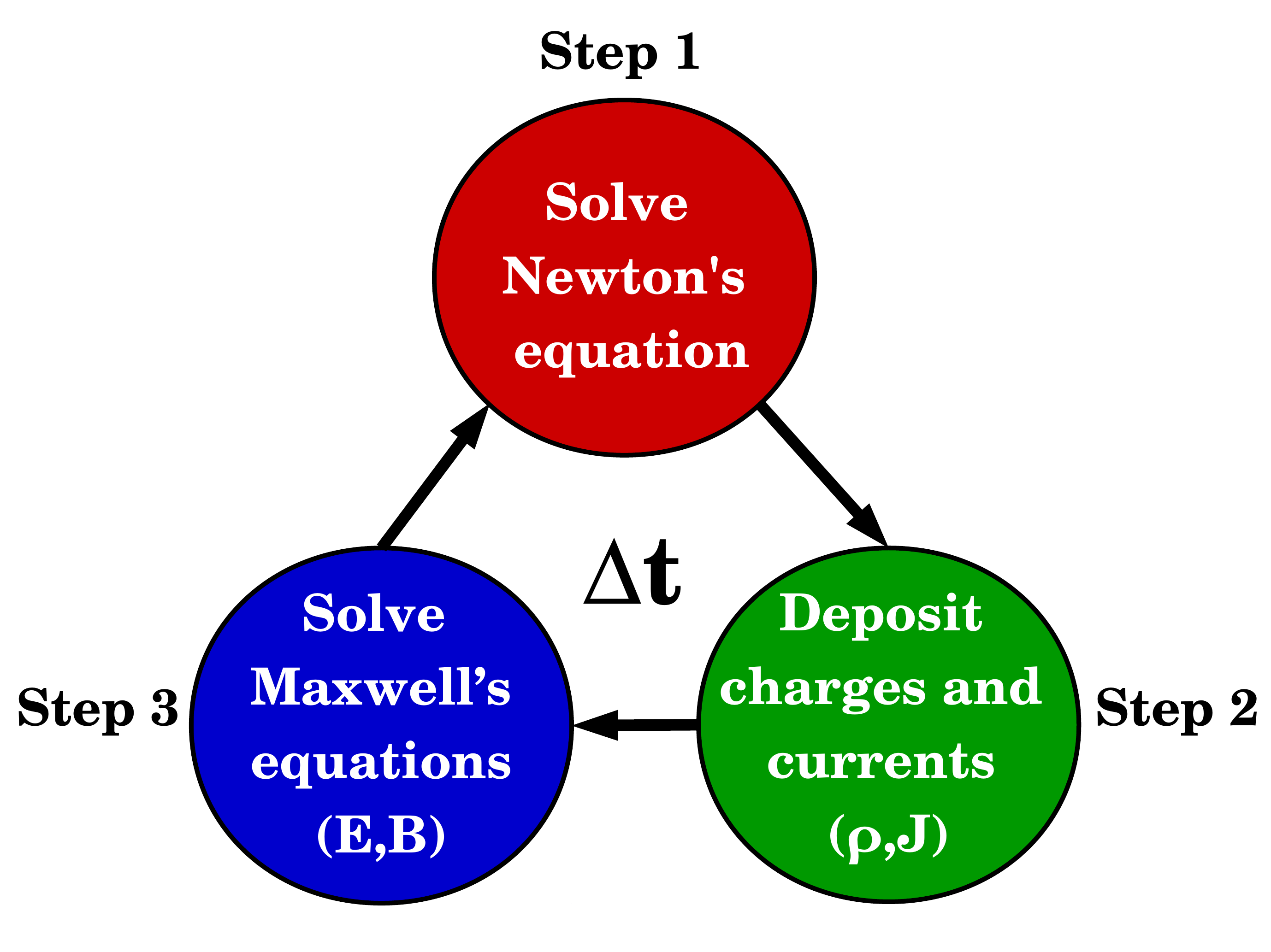}
\caption{Computation procedure per time step $\Delta t$ in PIC.}
\label{fig_loop}
\end{figure}

\subsubsection*{Step 1: Particle push}

\begin{figure}[]
\centering
\includegraphics[width=11.50cm]{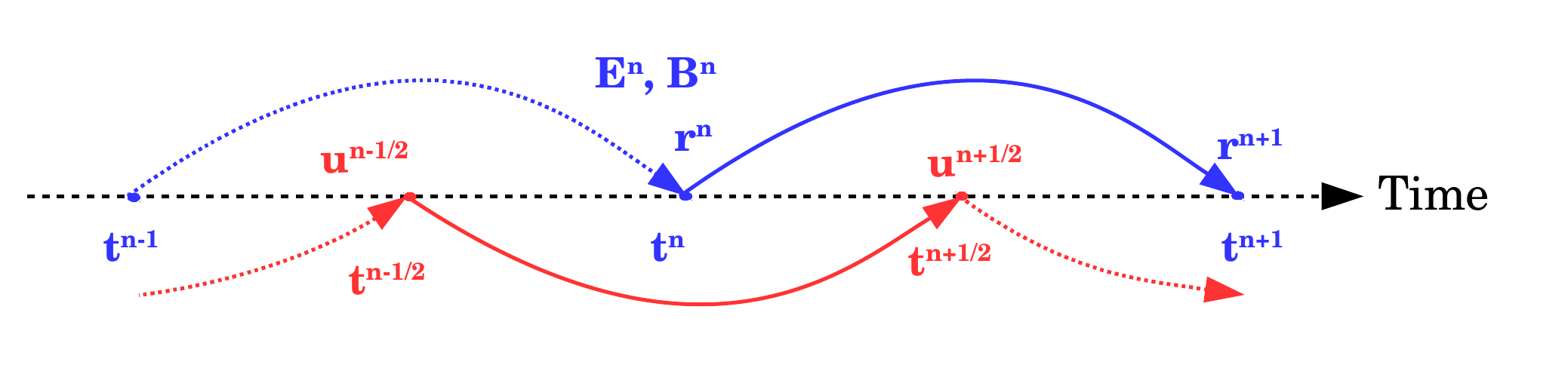}
\caption{The leapfrog scheme of the Boris method to solve Newton's equation.}
\label{fig_boris}
\end{figure}

The set of equations to solve are
\begin{eqnarray}
\frac{d\mathbf{u}}{dt} &=& \frac{q}{mc}\left(\mathbf{E}+ \frac{\mathbf{u} \times\mathbf{B}}{\gamma}\right)\label{eq_newton}\\
\frac{d\mathbf{r}}{dt} &=& \frac{c\mathbf{u}}{\gamma},
\end{eqnarray}
where $\mathbf{u}=\gamma\mathbf{v}/c$ is the particle 4-velocity vector divided by the speed of light and $\gamma=1/\sqrt{1-\left(v/c\right)^2}$ is the Lorentz factor. One of the most successful and most common method used in PIC to solve Newton's equation is the Boris push\footnote{Other efficient methods exists as for instance the particle pusher developed by \citet{2008PhPl...15e6701V}.}. It has all the desirable numerical features one might think of: it is fast, stable, second order accurate, and conserves well the particle energy. The algorithm is based on the usual leapfrog integration method, i.e., 4-velocities $\mathbf{u}$ and positions $\mathbf{r}$ are staggered in time by half a timestep (Figure~\ref{fig_boris}). If particle positions and fields are known at time $t^n$ ($\mathbf{r}^{n}$, $\mathbf{E}^{n}$, $\mathbf{B}^{n}$) and velocities at time $t^{n-1/2}$ ($\mathbf{u}^{n+1/2}$), the finite-difference time-centered expression of Eq.~(\ref{eq_newton}) is
\begin{equation}
\frac{\mathbf{u}^{n+1/2}-\mathbf{u}^{n-1/2}}{\Delta t}=\frac{q \mathbf{E}^{n}}{m c}+ \frac{q}{mc}\left(\frac{\mathbf{u}^{n}\times \mathbf{B}^{n}}{\gamma^n}\right).\label{eq_boris}
\end{equation}
Now, the trick is to rewrite $\mathbf{u}^{n}$ appearing on the right-hand side of the equation as $\mathbf{u}^{n}=\left(\mathbf{u}^{n-1/2}+\mathbf{u}^{n+1/2}\right)/2$. Assuming that $\mathbf{E}^{n}$ and $\mathbf{B}^{n}$ are known, $\mathbf{u}^{n+1/2}$ can be extracted. It is convenient to define the following intermediate variables
\begin{eqnarray}
\mathbf{u^{-}} &=& \mathbf{u}^{n-1/2}+\frac{q\Delta t\mathbf{E}^n}{2mc} \label{eq_u-}\\
\mathbf{u^{+}} &=& \mathbf{u}^{n+1/2}-\frac{q\Delta t\mathbf{E}^n}{2mc}\label{eq_u+}.
\end{eqnarray}
Then, using Eqs.~(\ref{eq_u-})-(\ref{eq_u+}) and after a few algebraic manipulations one finds
\begin{equation}
\mathbf{u^+} = \mathbf{u^-}+\mathbf{u^-}\times\mathbf{s}+\left(\mathbf{u^-}\times\mathbf{w}\right)\times \mathbf{s},
\end{equation}
where
\begin{equation}
\mathbf{w}=\frac{q\Delta t \mathbf{B}^n}{2mc\gamma^n},\hspace{0.5cm}\mathbf{s}=\frac{2\mathbf{w}}{1+\mathbf{w}^2},
\hspace{0.5cm}\gamma^{n}=\sqrt{1+\left(\mathbf{u}^-\right)^2}.
\end{equation}
It is important to notice that the fields appearing in these equations are those felt at the particle position, not at the grid point where the fields are known. The fields must be interpolated to the particle positions. A linear interpolation scheme is usually sufficient. The final step is to update the particle positions
\begin{equation}
\mathbf{r}^{n+1}=\mathbf{r}^{n}+c\Delta t\frac{\mathbf{u}^{n+1/2}}{\gamma^{n+1/2}},
\end{equation}
where $\gamma^{n+1/2}=\sqrt{1+\left(\mathbf{u}^{n+1/2}\right)^2}$.

\subsubsection*{Step 2: Charge and current deposition}

To solve Maxwell's equations, we need the source terms $\rho$ and $\mathbf{J}$ that are given by the particles. In a continuous space, these macroscopic quantities can be recovered by summing over the contribution from all particles
\begin{equation}
\rho\left(\mathbf{r}\right)=\sum_{k=1}^{N}q_k w_k \delta\left(\mathbf{r}-\mathbf{r}_k\right), \hspace{5mm} \mathbf{J}\left(\mathbf{r}\right)=\sum_{k=1}^{N}q_k w_k \mathbf{v}_k\delta\left(\mathbf{r}-\mathbf{r}_k\right),
\end{equation}
where $q_k$, $\mathbf{v}_k$ are respectively the electric charge and the 3-velocity of the particle $k$. In PIC, charges and currents from the particles must be collected and dispatched among the nearest grid points. Charge and current densities at the grid point $\mathbf{r}_{i}$ can be written as
\begin{equation}
\mathbf{\rho}\left(\mathbf{r}_i\right)=\sum_{k=1}^{N}q_k w_k S\left(\mathbf{r}_i-\mathbf{r}_k\right), \hspace{5mm} \mathbf{J}\left(\mathbf{r}_i\right)=\sum_{k=1}^{N}q_k w_k \mathbf{v}_k S\left(\mathbf{r}_i-\mathbf{r}_k\right),
\end{equation}
where $S$ is a shape function which depends on the desired deposition scheme. Even though the particles are point-­like, they have an effective size that is felt through the deposition of currents on the grid.

\begin{figure}[]
\centering
\includegraphics[width=7cm]{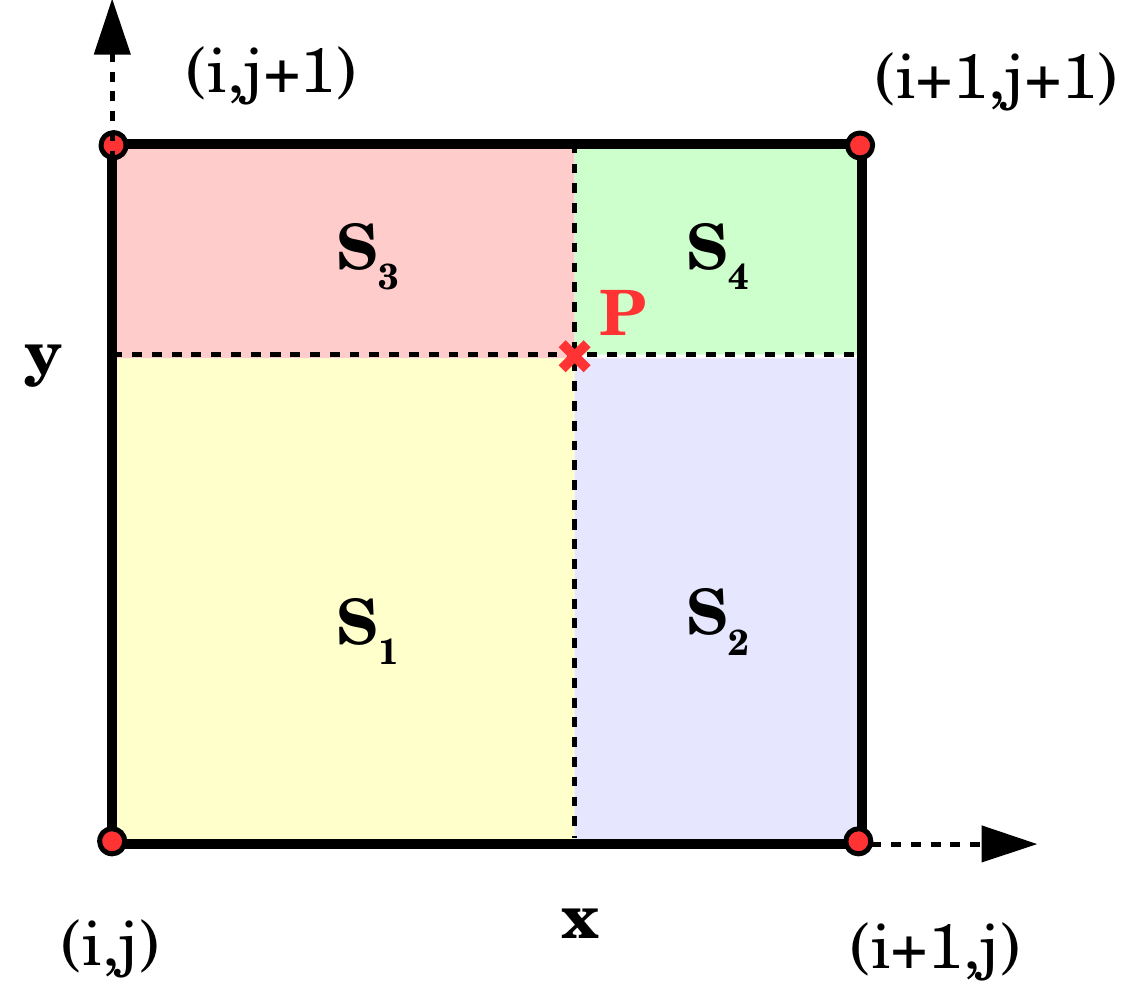}
\caption{The area-weighting technique to interpolate fields or deposit charges and currents on a 2D Cartesian cell $(x_i,y_i)$ for a particle located in $P(x,y)$. The contribution to node $(i,j)$ is given by $S_4/S_{\rm tot}$, to $(i+1,j)$ is $S_3/S_{\rm tot}$, to $(i,j+1)$ is $S_2/S_{\rm tot}$ and to $(i+1,j+1)$ is $S_1/S_{\rm tot}$.}
\label{fig_deposit}
\end{figure}

Figure~\ref{fig_deposit} shows the example of a first order linear deposition method in a 2D Cartesian grid cell (or area-weighting method). The contributions from all the particles contained in the cell $(x_i,y_j)$ to the current $\mathbf{J}$ are given by
\begin{eqnarray}
\mathbf{J}_{i,j} &=& \sum_{k=1}^{N_{\rm cell}}q_k w_k \mathbf{v}_k \left(1-a_{ k}\right)\left(1-b_{k}\right)\\
\mathbf{J}_{i+1,j} &=& \sum_{k=1}^{N_{\rm cell}}q_k w_k \mathbf{v}_k a_{ k}\left(1-b_{k}\right)\\
\mathbf{J}_{i,j+1} &=& \sum_{k=1}^{N_{\rm cell}}q_k w_k \mathbf{v}_k \left(1-a_{k}\right)b_{k}\\
\mathbf{J}_{i+1,j+1} &=& \sum_{k=1}^{N_{\rm cell}}q_k w_k \mathbf{v}_k a_{k} b_{k},
\end{eqnarray}
where
\begin{equation}
a_{k}=\frac{x_{k}-x_{i}}{dx},\hspace{5mm} b_{k}=\frac{y_{k}-y_{j}}{dy}
\end{equation}
are the usual bilinear interpolation coefficients. In this particular example, particles have a triangular shape.

\subsubsection*{Step 3: Fields evolution}

The last step is to update the fields on the grid. In principle, one needs solely to solve the time-dependent equations
\begin{eqnarray}
\frac{\partial \mathbf{E}}{\partial t} &=& c\boldsymbol{\nabla}\times\mathbf{B}-4\pi\mathbf{J}\\
\frac{\partial \mathbf{B}}{\partial t} &=& -c\boldsymbol{\nabla}\times\mathbf{E},\label{eq_maxb}
\end{eqnarray}
because the current density is already given by the particles as we have seen in the previous paragraph. The other two should be automatically satisfied, but this is not necessarily true due to truncation errors in the discretization of space and time derivatives. The current deposition procedure does not always guarantee charge conservation\footnote{The total particle charge is conserved, but not necessarily the charge deposited on the grid.} (i.e., $\boldsymbol{\nabla}\cdot\mathbf{J}\neq-\partial\rho/\partial t$), but some solutions exist to enforce it to machine roundoff precision \citep{1992CoPhC..69..306V, 2001CoPhC.135..144E}. Alternatively, Poisson equation should be solved to correct the electric field to make sure $\boldsymbol{\nabla}\cdot\mathbf{E}=4\pi\rho$. Parabolic and hyberbolic divergence cleaning methods also exist in the literature \citep{1987JCoPh..68...48M, 2000JCoPh.161..484M}.

\begin{figure}[]
\centering
\includegraphics[width=11.50cm]{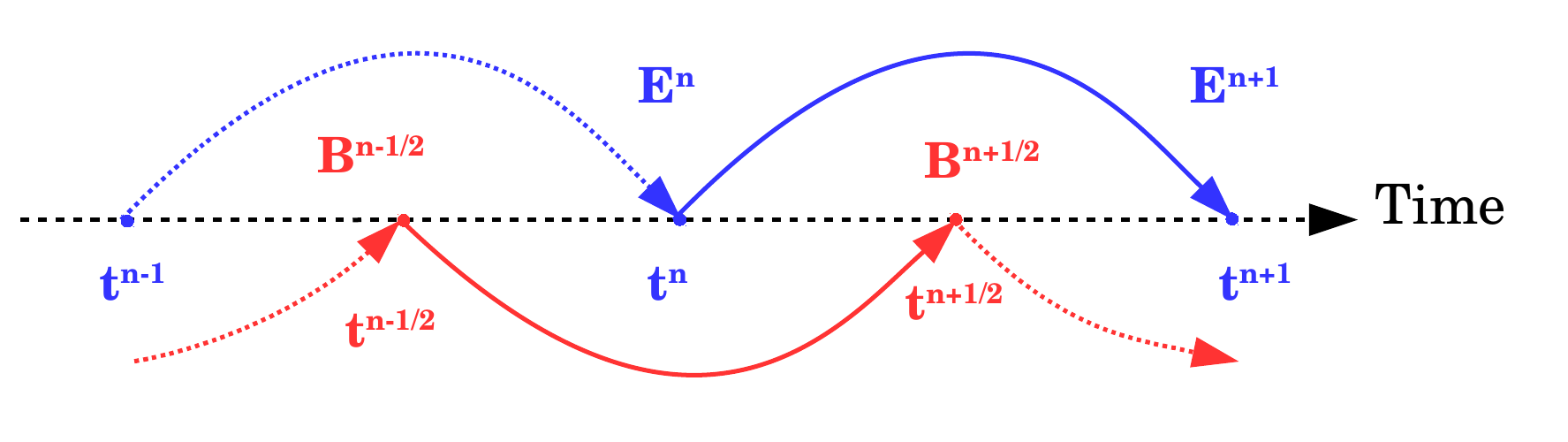}
\includegraphics[width=5.75cm]{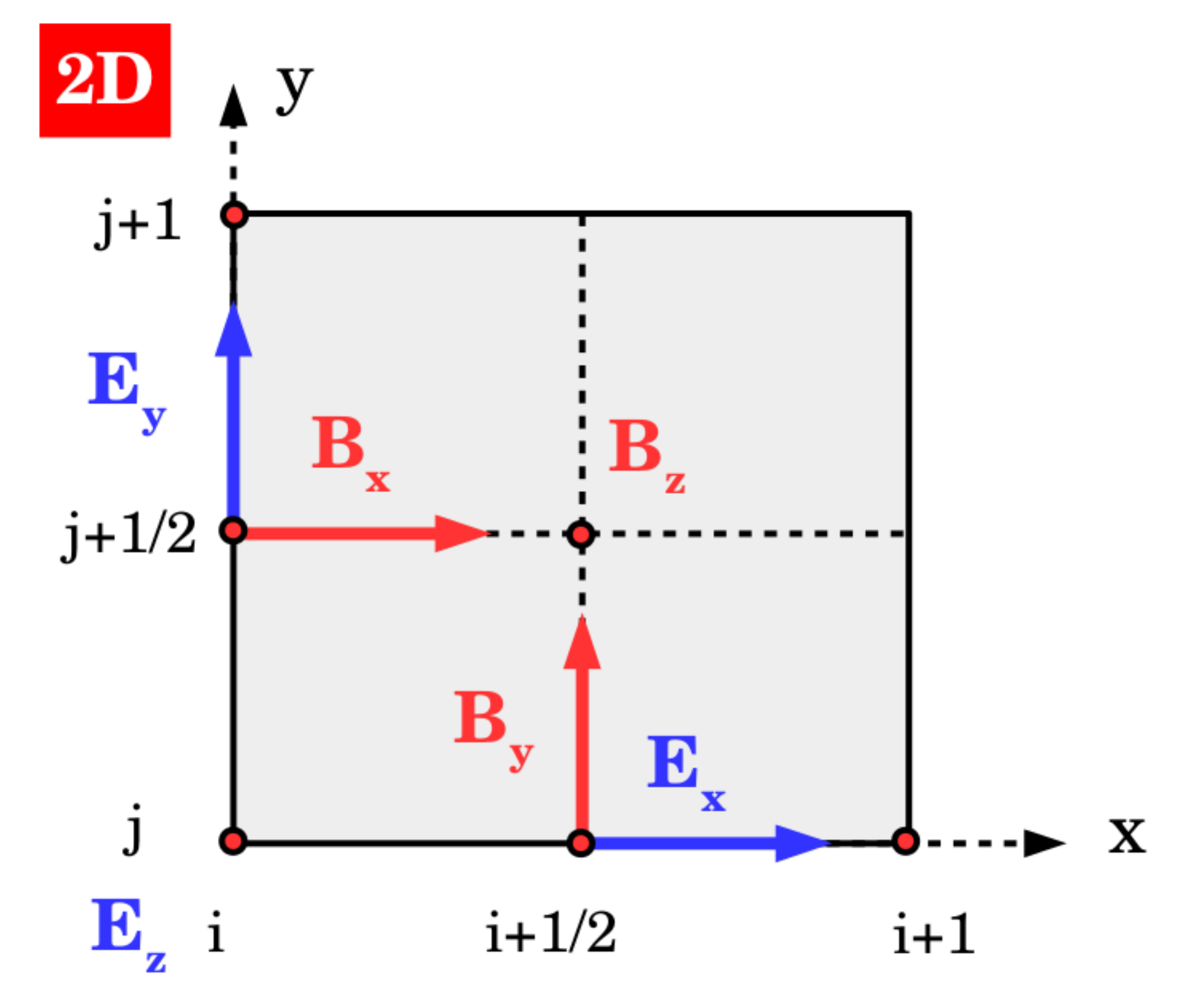}
\includegraphics[width=5.75cm]{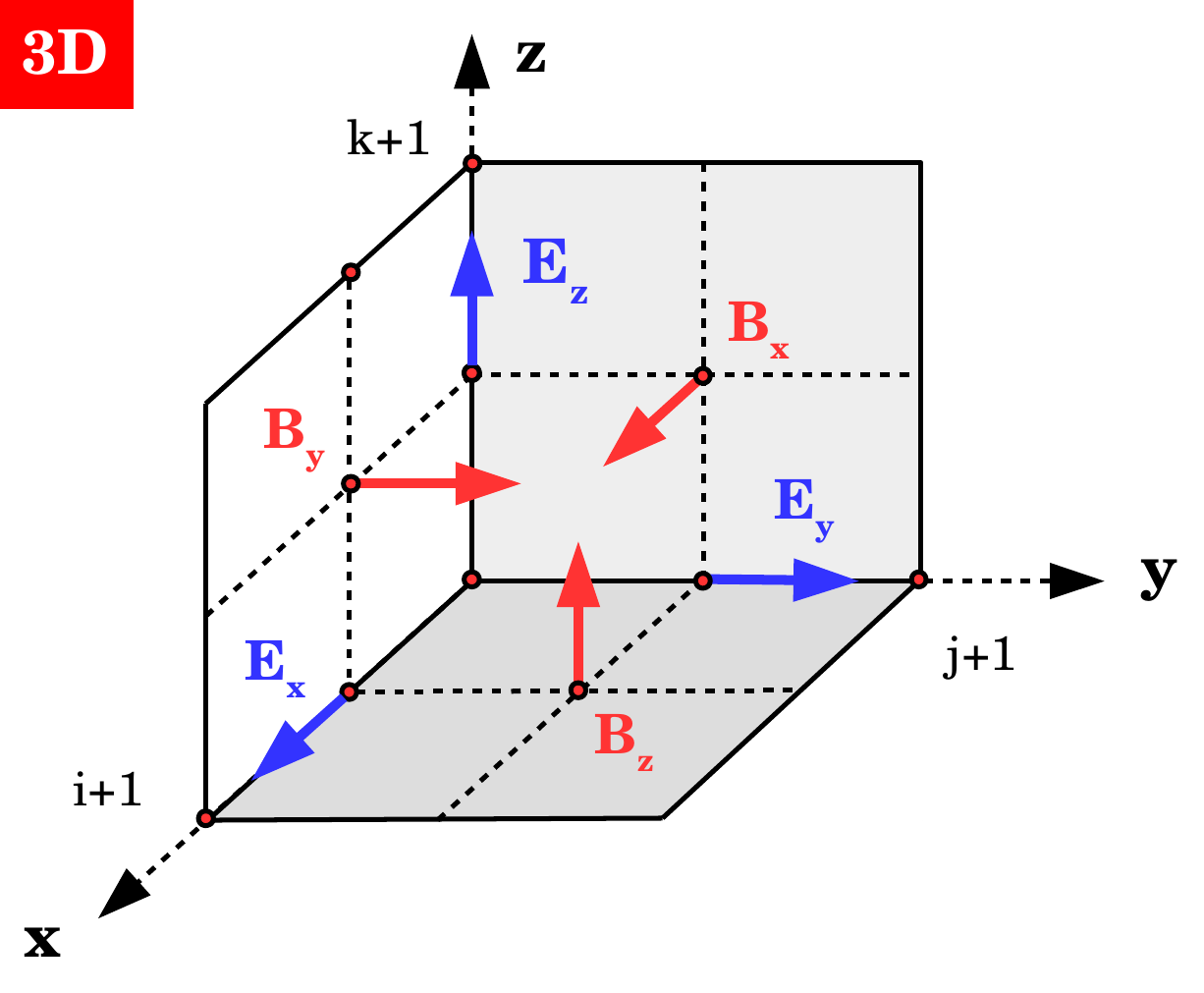}
\caption{Top: Leapfrog scheme for the fields in time. Bottom: Staggered mesh proposed by \citet{1966ITAP...14..302Y} in 2D (left) and in 3D (right).}
\label{fig_yee}
\end{figure}

The finite difference time domain (FDTD) method proposed by \citet{1966ITAP...14..302Y} for solving the time-dependent Maxwell equations enforces $\boldsymbol{\nabla}\times\mathbf{B}=0$ to machine roundoff precision. This is the most commonly used method in explicit PIC codes. Like the Boris push, the FDTD method combines stability, efficiency and second order accuracy (here in both space and time). To achieve this, fields must be staggered in time and in space. Figure~\ref{fig_yee} shows the order in time (top panel), as well as the spatial configuration of the fields within a Cartesian cell\footnote{For a spherical geometry, see \citealt{1983ITNS...30.4592H, 2015MNRAS.448..606C, 2016MNRAS.457.2401C, 2015NewA...36...37B}.} in 2D (bottom-left panel) and in 3D (bottom-right panel). For illustrative purposes, within this framework the $z$-component of Eq.~(\ref{eq_maxb}) is
\begin{eqnarray}
\frac{\left(B_{\rm z}\right)^{n+1/2}_{i+1/2,j+1/2,k}-\left(B_{\rm z}\right)^{n-1/2}_{i+1/2,j+1/2,k}}{\Delta t}=-c \frac{\left(E_{\rm y}\right)^{n}_{i+1,j+1/2,k}-\left(E_{\rm y}\right)^{n}_{i,j+1/2,k}}{\Delta x} \nonumber \\
+c \frac{\left(E_{\rm x}\right)^{n}_{i+1/2,j+1,k}-\left(E_{\rm x}\right)^{n}_{i+1/2,j,k}}{\Delta y},
\end{eqnarray}
where $\Delta x,~\Delta y$ are the spatial step size along $x$ and $y$. The FDTD method is stable under the usual Courant-­Friedrichs-­Lewy condition, i.e.,
\begin{eqnarray}
\left(\frac{c\Delta t}{\Delta x}\right)^2~\rm{(1D)} &<& 1,\\ \nonumber
\left(c\Delta t\right)^2\left(\frac{1}{\Delta x^2}+\frac{1}{\Delta y^2}\right) &<& 1~\rm{(2D)},\\ \nonumber
\left(c\Delta t\right)^2\left(\frac{1}{\Delta x^2}+\frac{1}{\Delta y^2}+\frac{1}{\Delta z^2}\right) &<& 1~\rm{(3D)}.
\end{eqnarray}
This is a purely numerical requirement, but physics imposes other constraints on the size of the steps, namely that the Debye length and the plasma frequency are well resolved by the code ($\Delta x/\Lambda_{\rm D}\ll 1$ and $\omega_{\rm pe}\Delta t\ll 1$), the latter condition being more stringent.

\subsection{Boundary conditions}

Periodic boundary conditions are robust, easy to implement and physically useful in many case studies but there are not always appropriate. Below is a brief description of some other boundary conditions usually employed in PIC simulations.

\subsubsection*{Particles}

\begin{itemize}
\item{A perfectly reflective wall scatters the particle with no loss of momentum and energy. It can be useful in the context of a perfectly conducting wall for the fields (see below).}
\item{In the context of an absorbing wall or of an open boundary, it is appropriate to absorb particles at the boundary. In such a case, particles are simply removed from the simulation.}
\item{Conversely, new particles can be injected in simulations. This can be motivated by the physics involved in the problem, as for instance in pulsars where pair creation is important \citep{2013MNRAS.429...20T, 2014ApJ...795L..22C, 2015ApJ...801L..19P}, or by the numerical setup if for example there is an inflow from one side of the box as in PIC simulations of relativistic shocks \citep{spitkovsky_08b, sironi_spitkovsky_11b}. In this case, an expanding box with an injector receding at the speed of light can be desirable to reduce numerical cost (see \ref{sect_applications} below).}
\end{itemize}

\subsubsection*{Fields}

\begin{itemize}
\item{Perfectly conducting walls allow to reflect electromagnetic waves. They are easily implemented by applying the usual boundary conditions, namely that the tangential component of $\mathbf{E}$ and the perpendicular component of $\mathbf{B}$ vanishes at the interface. Semi-reflective medium can also be easily coded using surface current and charge densities.}
\item{It is sometimes useful to absorb all electromagnetic waves leaving the box to simulate an open boundary, as for instance in pulsar winds \citep{2015MNRAS.448..606C, 2015NewA...36...37B}. In this case, the open boundary is coated with an absorbing medium of several cell thick where resistive terms are added to Maxwell's equations
\begin{eqnarray}
\frac{\partial \mathbf{E}}{\partial t}+{\color{red}\lambda\mathbf{E}} &=& c\boldsymbol{\nabla}\times\mathbf{B}-4\pi\mathbf{J}\\ \label{eq_pml1}
\frac{\partial \mathbf{B}}{\partial t} +{\color{red}\lambda^{\star}\mathbf{B}}&=& -c\boldsymbol{\nabla}\times\mathbf{E},\label{eq_pml2}
\end{eqnarray}
where $\lambda$ and $\lambda^{\star}$ are artificial electric and magnetic ``conductivities''. The transition between the working domain and the absorbing layer should be gradual to avoid undesired reflections at the boundary. Conductivities usually are increasing function from the inner edge to the outer edge of the damping layer to make sure waves are completely absorbed. Eqs.~(\ref{eq_pml1}-\ref{eq_pml2}) are valid for 1D layer. A perfectly matched layer is a generalization of these formulae to a multidimensional damping layer \citep{1994JCoPh.114..185B, 1996JCoPh.127..363B}. In this framework, fields must be split into two subcomponents and hence the number of equations to solve is doubled (up to 12 in 3D, for an application to pulsars see e.g., \citealt{2009A&A...496..495K} in the context of force-free MHD simulations).}
\end{itemize}

\subsection{Parallelization}

\begin{figure}[]
\centering
\includegraphics[width=10cm]{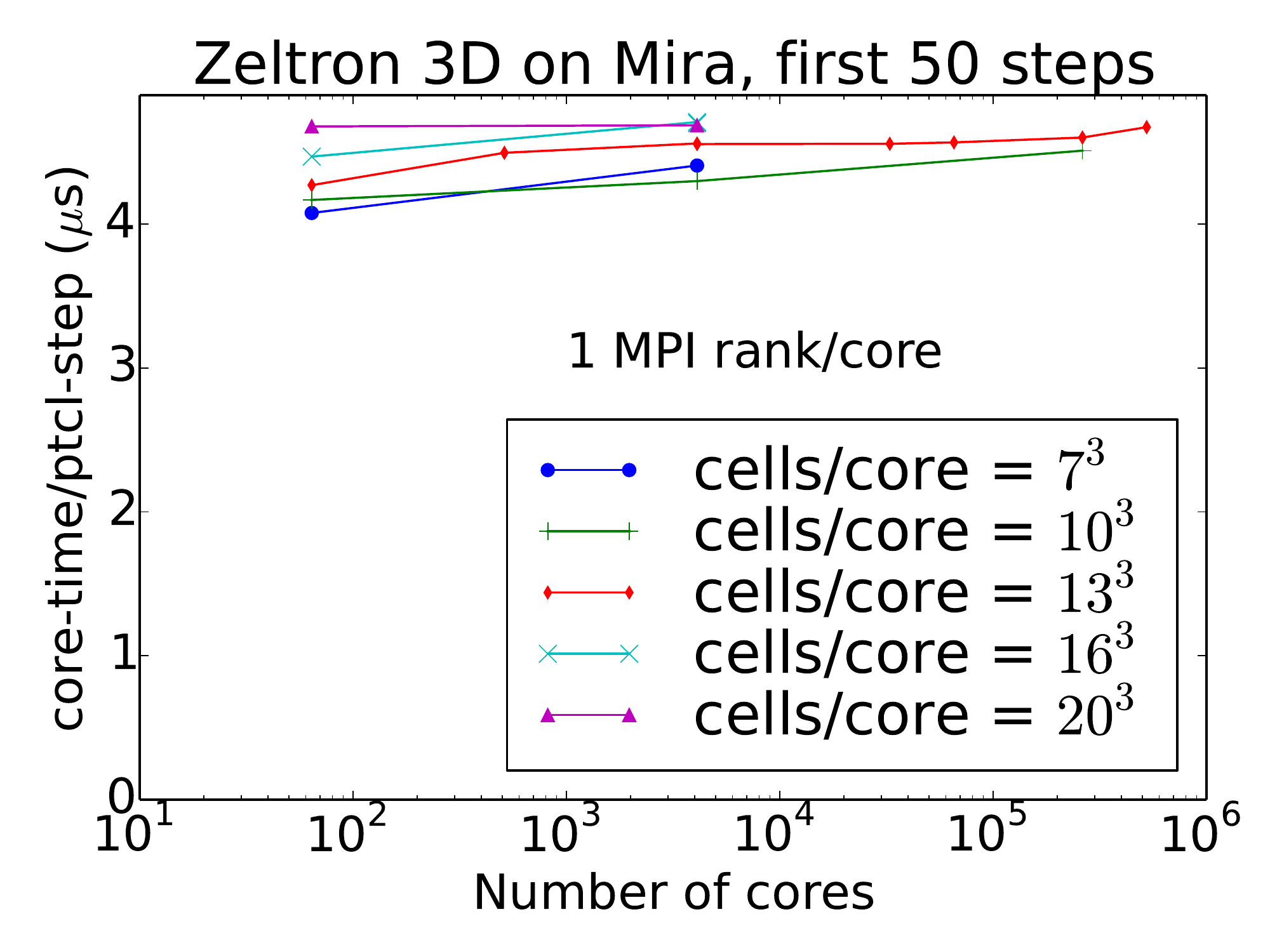}
\caption{Parallel scaling performance of the {\tt zeltron} PIC code on the Mira supercomputer at the Argonne Leadership Computing Facility. Courtesy Greg Werner.}
\label{fig_performance}
\end{figure}

PIC codes must be efficiently parallelized to model large system size and long integration time to have meaningful astrophysical applications. A common practice is to use the domain decomposition technique. It consists in dividing the computational box into smaller domains where one or more cores are assigned. Each CPU goes through the main steps described in Sect.~\ref{sect_loop} every timestep and exchanges information with the neighbouring processes to send particle and field data at the interface between subdomains. Communications between an arbitrary number of processes are done thanks to the Message Passing Interface (MPI) library. PIC codes scale well to a large number of CPUs, today at least up to $\sim 10^6$ processes (see Fig.~\ref{fig_performance}). These scaling plots are usually done under ideal conditions, and do not necessarily reflect problem-dependent loss of performance. In PIC, a poor load balancing severely slows down a simulation. If, for some reason, there is a concentration of particle in a few subdomains, only few processors will have to push a lot of particles while the others remain idle\footnote{Typically, pushing particles and depositing currents take $90\%$ of the computing time (without communications), this is the reason why load-balancing is so critical in PIC.}. The way how the domain is decomposed for a given setup can usual make a big difference. Hybrid codes combining MPI and OpenMP, variable particle weighting, or dynamical changes of the domain decomposition are other solutions to have better performances.

%%%%%%%%%%%%%%%%%%%%%%%%%%%
\section{Application to pulsar wind nebulae}\label{sect_applications}
In this section, we describe the main results obtained with PIC simulations on the efficiency of particle acceleration in PWNe. This section is divided into two parts: at first, we investigate particle acceleration at the termination shock of the pulsar wind, which is usually invoked to power the nebular {\it quiescent} emission (\sect{sect:quie}); then, we will discuss the origin of the gamma-ray {\it flares} detected from the Crab Nebula in the last few years (\sect{sect:flare}), focusing on the mechanisms that might explain such extreme particle acceleration events.

\subsection{The quiescent emission}\label{sect:quie}
The Crab Nebula, our best laboratory for high energy astrophysics, has been observed over the entire electromagnetic spectrum from $\lesssim100\, {\rm MHz}$ to $\gtrsim1.5\, {\rm TeV}$ (see several contributions in this volume). Efficient acceleration of particles  at the termination shock  is required to explain its broadband spectrum. However, the flatness of the radio spectrum \citep[$F_{\nu_r}\propto\nu^{-0.3}$,][]{bietenholz_97} is hard to reconcile with the steeper optical and X-ray slope \citep[$F_{\nu_X}\propto\nu^{-1.1}$,][]{mori_04}, unless the electron distribution is more complicated than a single power law. In fact, the radio band would require a distribution of emitting particles with a power-law slope $p=-d \log N/d\log \gamma\simeq1.6$, whereas $p\gtrsim2$ would be needed for the optical and X-ray emission. Even more fundamentally, how the pulsar wind termination shock can accelerate particles to the required ``non-thermal'' energies (i.e., well beyond the ``thermal'' peak of a Maxwellian distribution) is still an unsolved problem.

Particle acceleration in shocks is usually attributed to the Fermi process, where particles are energized by bouncing back and forth across the shock. Despite its importance, the Fermi process is still not understood from first principles. The highly nonlinear coupling between accelerated particles and magnetic turbulence --- which is generated by the particles, and at the same time governs their acceleration --- is extremely hard to incorporate in analytic models, and can be captured only with {\it ab initio} PIC simulations \citep[for a review of the Fermi process in relativistic shocks, see][]{sironi_keshet_15}. 

As we describe below, the efficiency of the Fermi process depends critically on the shock properties, e.g., composition, magnetization (i.e., the ratio $\sigma$ between the Poynting flux and the kinetic energy flux of the pre-shock flow) and magnetic obliquity (i.e., the angle $\theta$ between the upstream magnetic field and the shock direction of propagation).\footnote{In the limit $\gamma_0\gg1$ of ultra-relativistic shocks, as appropriate for PWNe, the efficiency of the Fermi process does not depend on the shock Lorentz factor $\gamma_0$ \citep{sironi_spitkovsky_09,sironi_spitkovsky_11a,sironi_13}.}
Pulsar winds are thought to be dominated by electron-positron pairs \citep{bucciantini_11}. MHD models of PWNe require  $\sigma\gtrsim0.01-0.1$ in order to reproduce the morphology of the Crab jet/plume. Finally, polarization measurements indicate that the nebular magnetic field should be toroidal around the symmetry axis of the system, so that the  termination shock is ``perpendicular'' (i.e., with the field orthogonal to the flow direction).

PIC simulations of perpendicular magnetized shocks show negligible particle acceleration \citep{gallant_92,hoshino_08,sironi_spitkovsky_09,sironi_spitkovsky_11a,sironi_13}. Here, due to the lack of significant self-generated turbulence, charged particles are forced to slide along the background field lines, whose orientation prohibits repeated crossings of the shock. This inhibits the Fermi process, and in fact the particle distribution behind perpendicular shocks is purely thermal. 
%The same conclusion holds for both electron-positron and electron-ion flows. In electron-ion shocks, the incoming electrons are heated up to the ion energy, due to powerful electromagnetic waves emitted by the shock into the upstream medium, as a result of the synchrotron maser instability (studied analytically by \citet{lyubarsky_06}, and with 1D PIC simulations by e.g., \citet{langdon_88,gallant_92,hoshino_92,hoshino_08}). Yet, such heating is not powerful enough to permit an efficient injection of electrons into the Fermi acceleration process at perpendicular electron-ion shocks.

In summary, PIC simulations have shown that the shock configurations which are apparently most relevant for PWNe (i.e., ultra-relativistic magnetized perpendicular shocks) do not naturally result in efficient particle acceleration. This is in sharp contrast with the pronounced non-thermal signatures of the quiescent emission of PWNe. However, one key ingredient of the PIC results summarized above is that the pre-shock magnetic field direction stays {\it uniform} throughout the timespan of the simulations. This is generally not the case in pulsar winds. If the rotational and magnetic axes of the central pulsar are misaligned, around the equatorial plane the wind consists of toroidal stripes of {\it alternating} magnetic polarity, separated by current sheets of hot plasma. It is still a subject of active research whether the alternating stripes will dissipate their energy into particle heat ahead of the termination shock, or whether the wind remains dominated by Poynting flux till the termination shock \citep[][]{lyubarsky_kirk_01,kirk_sk_03,sironi_spitkovsky_11b}. If the stripes are dissipated far ahead of the termination shock, the upstream flow is weakly magnetized and the pulsar wind reaches a terminal Lorentz factor (in the frame of the nebula)
$
\gamma_0\sim L_{sd}/m_e c^2 \dot{N}\simeq3.7\times 10^{4} L_{sd,38.5}\dot{N}_{40}^{-1}~,
$
where $L_{sd}\equiv 3 \times 10^{38}L_{sd,38.5}\unit{erg s\,s^{-1}}$ is the spin-down luminosity of the Crab, and $\dot{N}=10^{40}\dot{N}_{40}\unit{s^{-1}}$ is the particle flux entering the nebula, including the radio-emitting electrons \citep{bucciantini_11}.

The two subsections below cover the two potential fates of the pulsar striped wind: at first, we investigate the physics of particle acceleration in a weakly magnetized shock (i.e., assuming that the alternating stripes have dissipated their magnetic energy far ahead of the termination shock); then, we assume that the stripes persist until the termination shock.

\subsubsection{The termination shock of a weakly magnetized wind}
\begin{figure}[!htb]
\begin{center}
\includegraphics[width=1.\textwidth,angle=0]{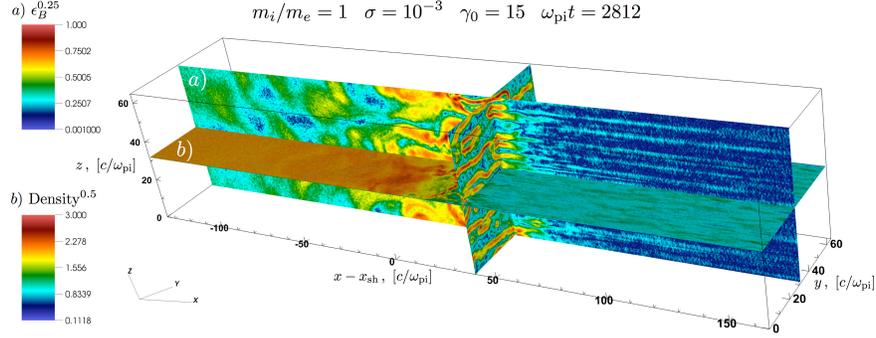}
\caption{\footnotesize{Shock structure from the 3D PIC simulation of a $\sigma=10^{-3}$ electron-positron shock with $\gamma_0=15$, from \cite{sironi_13}. The simulation is performed in the downstream frame and the shock propagates along $+\hat{x}$. We show the $xy$ slice of the particle number density (normalized to the upstream density), and the $xz$ and $yz$ slices of the magnetic energy fraction $\epsilon_B$. A stream of shock-accelerated particles propagates ahead of the shock, and their counter-streaming motion with respect to the incoming flow generates magnetic turbulence in the upstream via electromagnetic micro-instabilities. In turn, such waves provide the scattering required for particle acceleration.}}
\label{fig:shock}
\end{center}
\end{figure}

Weakly magnetized ultra-relativistic shocks are mediated by electromagnetic plasma instabilities \citep[the so-called Weibel instability,][]{weibel_59, medvedev_loeb_99, gruzinov_waxman_99}. These instabilities build up a magnetic barrier, up to a
level\footnote{The parameter $\epsilon_B$ denotes the magnetization of
  the turbulence, $\epsilon_B\,=\,\delta B^2/8\pi\gamma_0\rho_0c^2$, where $\delta B$ is the fluctuating magnetic field and $\rho_0$ is the mass density of the pre-shock flow. This should not be confused with the magnetization $\sigma=B_0^2/4\pi\gamma_0\rho_0c^2$, which quantifies the strength of the pre-existing ordered upstream field $B_0$.}
$\epsilon_B\,\sim\,10^{-2}-10^{-1}$, sufficient to deflect strongly
the incoming particles and thus mediate the shock transition. The instability --- triggered by a stream of shock-reflected particles propagating ahead of the shock --- generates filamentary magnetic structures in the upstream region (Fig.~\ref{fig:shock}), which in turn scatter the particles back and forth across the shock, mediating Fermi acceleration. 

Such shocks do self-consistently accelerate particles up to nonthermal energies, via the Fermi process \citep[][]{spitkovsky_08,spitkovsky_08b,martins_09,haugbolle_10,sironi_13}. The accelerated particles populate in the downstream region a power-law tail $dN/d\gamma\propto \gamma^{-p}$ with a slope $p\sim2.5$, that contains $\sim3\%$ of the particles and $\sim10\%$ of the flow energy.
%\footnote{These values are nearly independent of the flow composition and magnetization, in the regime of weakly magnetized shocks.} 

\begin{figure}[!hbt]
\begin{center}
\includegraphics[width=0.75\textwidth]{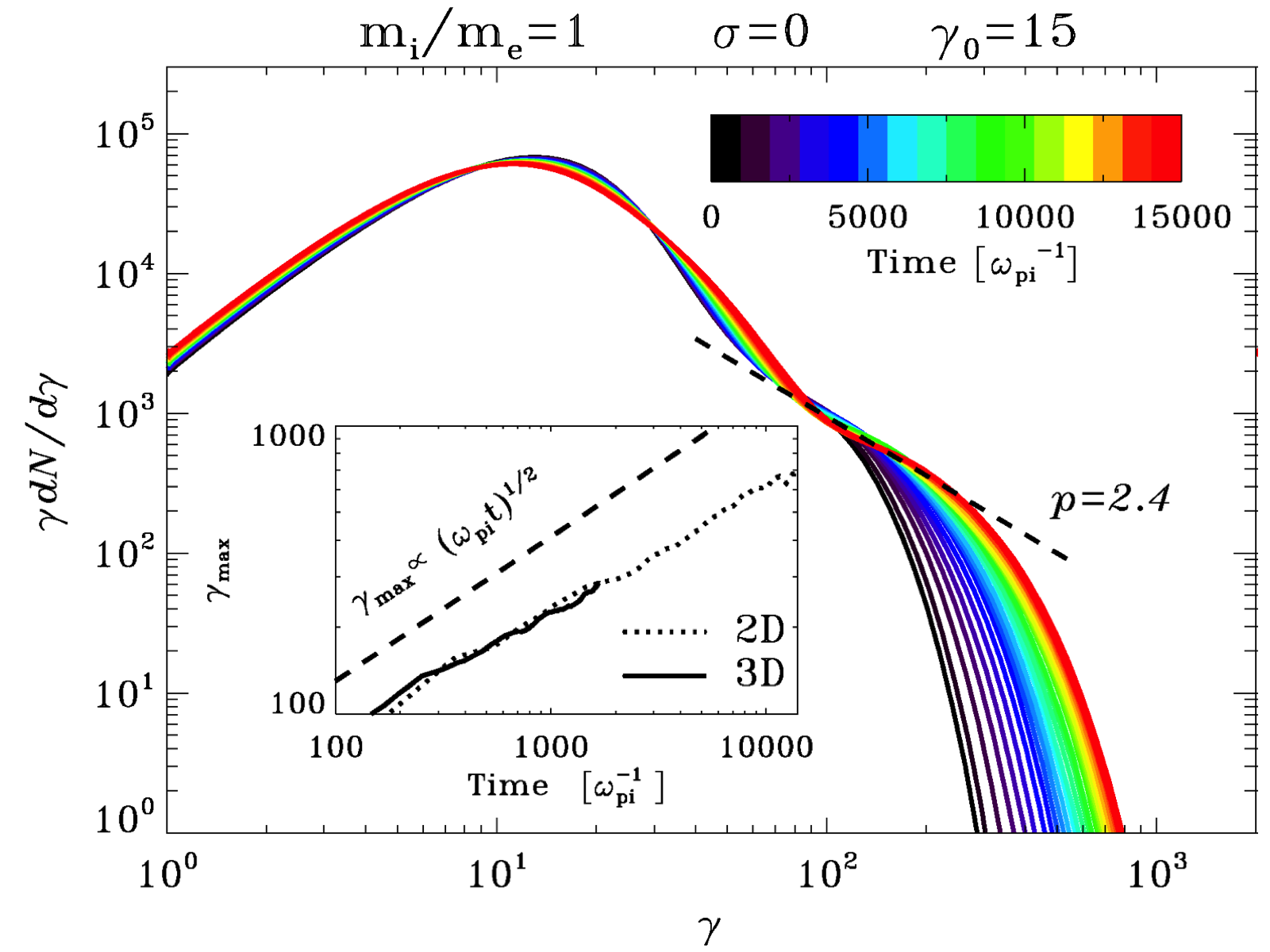}
\caption{\footnotesize{Temporal evolution of the downstream particle spectrum, from the 2D simulation of a $\gamma_0=15$ electron-positron shock propagating into an unmagnetized flow (i.e., $\sigma=0$), from \citet{sironi_13}. The evolution of the shock is followed from its birth (black curve) up to $\omega_{\rm p}t=15000$ (red curve). The non-thermal tail approaches at late times a power law with a slope $p\sim2.4$. Inset: temporal evolution of the maximum Lorentz factor, scaling as $\propto (\omega_{\rm p} t)^{1/2}$ (compare with the black dashed line) in both 2D (dotted) and 3D (solid).}}
\label{fig:spectime}
\end{center}
\end{figure}

The particle energy spectrum extends over time to higher and higher energies, as shown in  \fig{spectime}. For electron-positron flows, as appropriate for pulsar winds, the maximum post-shock particle Lorentz factor increases with time as $\gamma_{max}\sim 0.5 \,\gamma_0\, (\omega_{\rm p} t)^{1/2}$ \citep{sironi_13}.\footnote{This scaling is shallower than the so-called (and commonly assumed) Bohm limit $\gamma_{max}\propto t$, and it naturally results from the small-scale nature of the Weibel turbulence generated in the shock layer (see Fig.~\ref{fig:shock}).} The plasma frequency $\omega_{\rm p}$ can be computed from the number density ahead of the termination shock, which is $n_{{\rm TS}}=\dot{N}/(4 \pi R_{\rm TS}^2 c)$, assuming an isotropic particle flux. Here, $R_{\rm TS}\equiv3\times10^{17}R_{\rm TS,17.5}\unit{cm}$ is the termination shock radius. Balancing the acceleration rate with the synchrotron cooling rate in the self-generated Weibel fields, the maximum electron Lorentz factor is
\be
\gamma_{sync,e}\simeq3.5\times10^{8}L_{sd,38.5}^{1/6}\dot{N}_{40}^{-1/3} \epsilon_{B,-2.5}^{-1/3}R_{\rm TS,17.5}^{1/3}~.
\ee
A stronger constraint comes from the requirement that the diffusion length of the highest energy electrons be smaller than the termination shock radius (\ie a confinement constraint). Alternatively, the acceleration time should be shorter than $R_{\rm TS}/c$, which yields the critical limit
\be
\gamma_{\mathit{conf,e}}\simeq1.9\times10^{7}L_{sd,38.5}^{3/4}\dot{N}_{40}^{-1/2}~,
\ee
which is generally more constraining than the cooling-limited Lorentz factor $\gamma_{sync,e}$.
%This is the same constraint found by \citet{sironi_spitkovsky_11b} with PIC simulations of striped pulsar winds, even though they assumed that the magnetic stripes survive till the termination shock. In other words, we find that the limit on $\alpha$ for efficient Fermi acceleration does not depend on the location where the magnetic stripes dissipate (either in the wind, or at the shock).
The corresponding synchrotron photons will have energies
\be
\!\!\!h \nu_{\mathit{conf,e}}\simeq0.17\,L_{sd,38.5}^{2}\dot{N}_{40}^{-1}\epsilon_{B,-2.5}^{1/2}R_{\rm TS,17.5}^{-1}\unit{keV}
\ee
 which are apparently too small to explain the X-ray spectrum of the Crab, extending to energies beyond a few tens of MeV.
 
At face value, Fermi acceleration at the  termination shock of PWNe is not a likely candidate for producing X-ray photons via the synchrotron process.
Yet, the steady-state  hard X-ray and gamma-ray spectra of PWNe do look like the consequences of Fermi acceleration --- particle distributions with $p \simeq 2.4$ are a natural prediction of the Fermi process in ultra-relativistic shocks \citep{kirk_00,achterberg_01,keshet_waxman_05}. In this regard, we argue that the wind termination shock might form in a macroscopically turbulent medium, with the outer scale of the turbulence driven by the large-scale shear flows in the nebula \citep{komissarov_04,delzanna_04,camus_09}. If the large-scale motions drive a turbulent cascade to shorter wavelengths, back-scattering of the particles in this downstream turbulence, along with upstream reflection by the transverse magnetic field of the wind, might sustain Fermi acceleration to higher energies.
%Yet, the turbulent cascade has to work in such a way as not to disturb the polarization of the nebula, which looks rather cleanly toroidal \citep{wilson_72,schmidt_79}.

An alternative mechanism leading to particle acceleration to higher energies may be connected to the accelerator behind the recently discovered gamma-ray flares in the Crab Nebula (see Sect.~\ref{sect:flare}).
%If the stripes decay well ahead of the shock, the wind has a ``Mexican hat'' magnetic geometry, with oppositely wound toroidal magnetic field in the northern and southern hemispheres, separated by the equatorial current sheet. Tearing of that current sheet can create radial spokes of current, with radially extended X-lines.
Runaway acceleration of electrons and positrons at reconnection X-lines, a linear accelerator, may inject energetic beams into the shock, with the mean energy per particle approaching the whole open field line voltage, $\gtrsim 10^{16}\unit{V}$ in the Crab \citep{arons_12}, as required to explain the Crab GeV flares.  This high-energy population can drive cyclotron turbulence when gyrating in the shock-compressed fields, and resonant absorption of the cyclotron harmonics can accelerate the electron-positron pairs in a broad spectrum, with maximum energy again comparable to the whole open field line voltage \citep{hoshino_92,amato_arons_06}.

\subsubsection{The termination shock of a strongly magnetized striped wind}
\begin{figure}[tbp]
\begin{center}
\includegraphics[width=0.75\textwidth]{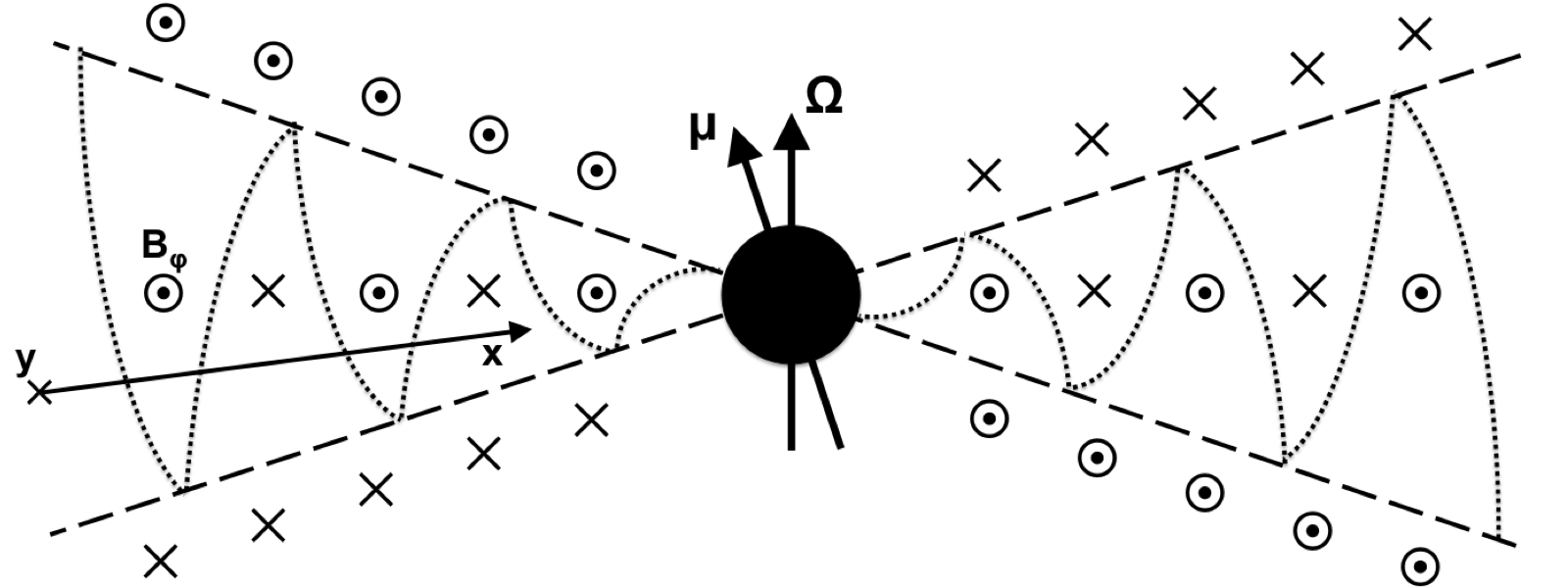}
\caption{Poloidal structure of the striped pulsar wind. The arrows denote the pulsar rotational axis (along $\bmath{\Omega}$, vertical) and magnetic axis (along $\bmath{\mu}$, inclined). Within the equatorial wedge bounded by the dashed lines, the wind consists of toroidal stripes of alternating polarity (see the reversals of $B_\phi$), separated by current sheets (dotted lines). At latitudes higher than the inclination angle between $\bmath{\Omega}$ and $\bmath{\mu}$ (i.e., beyond the dashed lines), the field does not alternate.The simulation domain is in the $xy$ plane, oriented as indicated.}
\label{fig:simplane}
\end{center}
\end{figure}

Assuming that the stripes survive until the termination shock, we now describe the physics of particle acceleration if the pre-shock flow carries a strong magnetic field of intensity $B_0$, oriented perpendicular to the shock direction of propagation and alternating with wavelength $\lambda$.\footnote{The wavelength $\lambda$ of the striped wind equals $c\,P$, where $P$ is the pulsar period.} Although the magnetic field strength in the  wind is always $B_0$, the wavelength-averaged field $\langle B_\phi\rangle_\lambda$ can vary from zero up to $B_0$, depending on the relative widths of the regions of  positive and negative field (see the sketch in \fig{simplane}). In  pulsar winds, one expects $\langle B_\phi\rangle_\lambda=0$ only in the equatorial plane (where the stripes are symmetric), whereas $|\langle B_\phi\rangle_\lambda|/B_0\rightarrow1$ at high latitudes. As a proxy for latitude, we choose $\alpha=2\langle B_\phi\rangle_\lambda/(B_0+|\langle B_\phi\rangle_\lambda|)$, which varies between zero and unity.

\begin{figure}[!tbp]
\centering 
\includegraphics[width=0.92\textwidth]{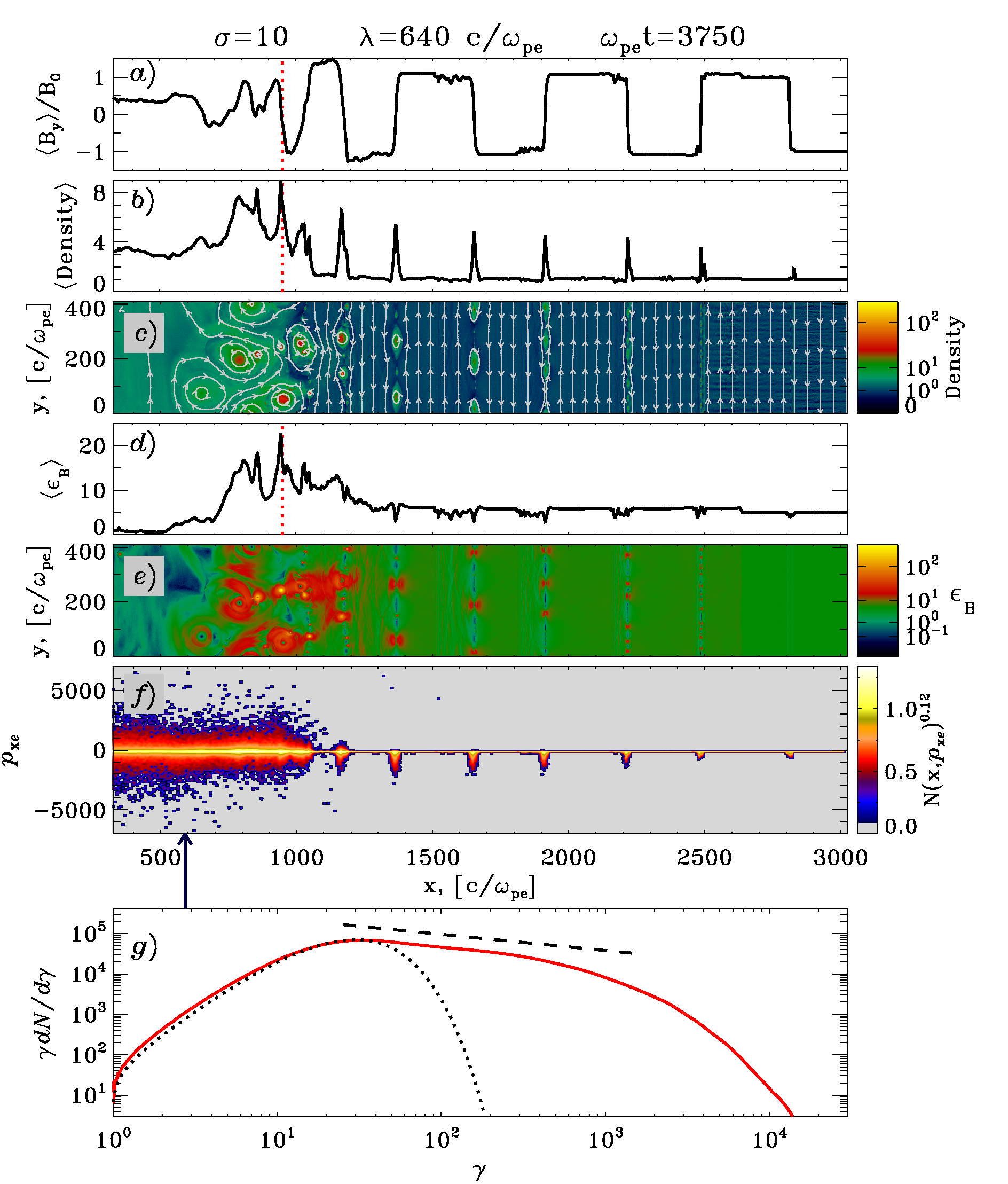}
\caption{\footnotesize{2D PIC simulation of a relativistic shock propagating in a striped flow with magnetization $\sigma=10$, $\alpha=0.1$ and stripe wavelength $\lambda=640\comp$, where $\comp$ is the so-called plasma skin depth, from \citet{sironi_spitkovsky_11b}. The shock is located at  $x\sim 950\,c/\omega_{\rm pe}$ (vertical dotted red line), and the incoming flow moves from right to left. At the shock, the striped structure of the  magnetic field  is erased (panel (a)), the flow compresses (density in panel (b)), and the field energy (panel (d)) is transferred to the particles (phase space  in panel (f)). The micro-physics of magnetic reconnection is revealed by the islands seen in the 2D plots of density and magnetic energy (panels (c) and (e), respectively) in a region around the shock. As a result of magnetic reconnection, the post-shock particle spectrum (red line in panel (g)) is much broader than a thermal distribution (dotted line), and it approaches a power-law tail with hard slope $p\simeq1.5$ (dashed line).}}
\label{fig:fluid}
\end{figure}

At the termination shock, the compression of the flow forces the annihilation of nearby field lines, a process known as driven magnetic reconnection \citep{lyubarsky_03, sironi_spitkovsky_11b,sironi_spitkovsky_12}. As shown in Fig.~\ref{fig:fluid}, magnetic reconnection erases the striped structure of the flow (panel (a)), and transfers most of the energy stored in the magnetic fields (panel (d)) to the particles, whose distribution becomes much hotter behind the shock (see panel (f), for $x\lesssim1000\comp$). As a result of field dissipation, the average particle energy increases by a factor of $\sigma$ across the shock, regardless of the stripe width $\lambda$ or the wind magnetization $\sigma$ (as long as $\alpha\lesssim0.1$). The reconnection process manifests itself as characteristic islands in density (panel (c)) and magnetic energy (panel (e)), separated by X-points where the magnetic field lines tear and reconnect.  

The incoming particles are accelerated by the reconnection electric field at the X-points and, in the post-shock spectrum, they populate a broad distribution (red line in panel (g)),  extending to much higher energies than  expected in thermal equilibrium (dotted line). For the parameters studied in \fig{fluid}, the slope of the non-thermal tail is $p\simeq1.5$ (dashed line in panel (g)), harder than what  the Fermi process normally gives in relativistic shocks.\footnote{Hard particle spectra are found to be a generic by-product of magnetic reconnection in the relativistic regime appropriate for pulsar winds \citep[e.g.,][]{sironi_spitkovsky_14,Guo:2014aa,melzani14,Werner:2016aa,sironi_15,sironi_16}.} While efficient field dissipation (and so, efficient transfer of field energy to the particles) occurs irrespective of the wind properties (if $\alpha\lesssim 0.1$), the width of the downstream particle spectrum is sensitive to the stripe wavelength and the wind magnetization through the combination $\lambda/r_{L,hot}$, namely the stripe wavelength measured in units of the {post-shock} particle Larmor radius (i.e., after dissipation has taken place, and the mean particle energy has increased by a factor of $\sigma$). A Maxwellian-like spectrum is obtained for $\lambda/r_{L,hot}\lesssim$ a few tens, whereas in the limit $\lambda/r_{L,hot}\gg1$ the spectrum approaches a broad power-law tail of index $1<p<2$, extending from $\gamma_{\rm min}\simeq\gamma_0$ up to $\gamma_{\rm max}\simeq\gamma_0\sigma^{1/(2-p)}$.

The particles are accelerated primarily by the reconnection electric field at the X-points, rather than by bouncing back and forth across the shock, as in the standard Fermi mechanism. Quite surprisingly, the Fermi process can still operate along the equatorial plane of the wind, where the stripes are quasi-symmetric ($\alpha\lesssim0.01$). Here, the highest energy particles accelerated by the reconnection electric field can escape ahead of the shock, and be injected into a Fermi-like acceleration cycle. In the post-shock spectrum, they populate a power-law tail with slope $p\simeq2.5$, that extends beyond the hard component produced by reconnection.
%\footnote{This additional population of Fermi-accelerated particles is not present in panel (g) of \fig{fluid}, since the figure focuses on the characteristic shock structure at intermediate latitudes  away from  the wind midplane.}

\begin{figure}[tbp]
\begin{center}
\includegraphics[width=0.75\textwidth]{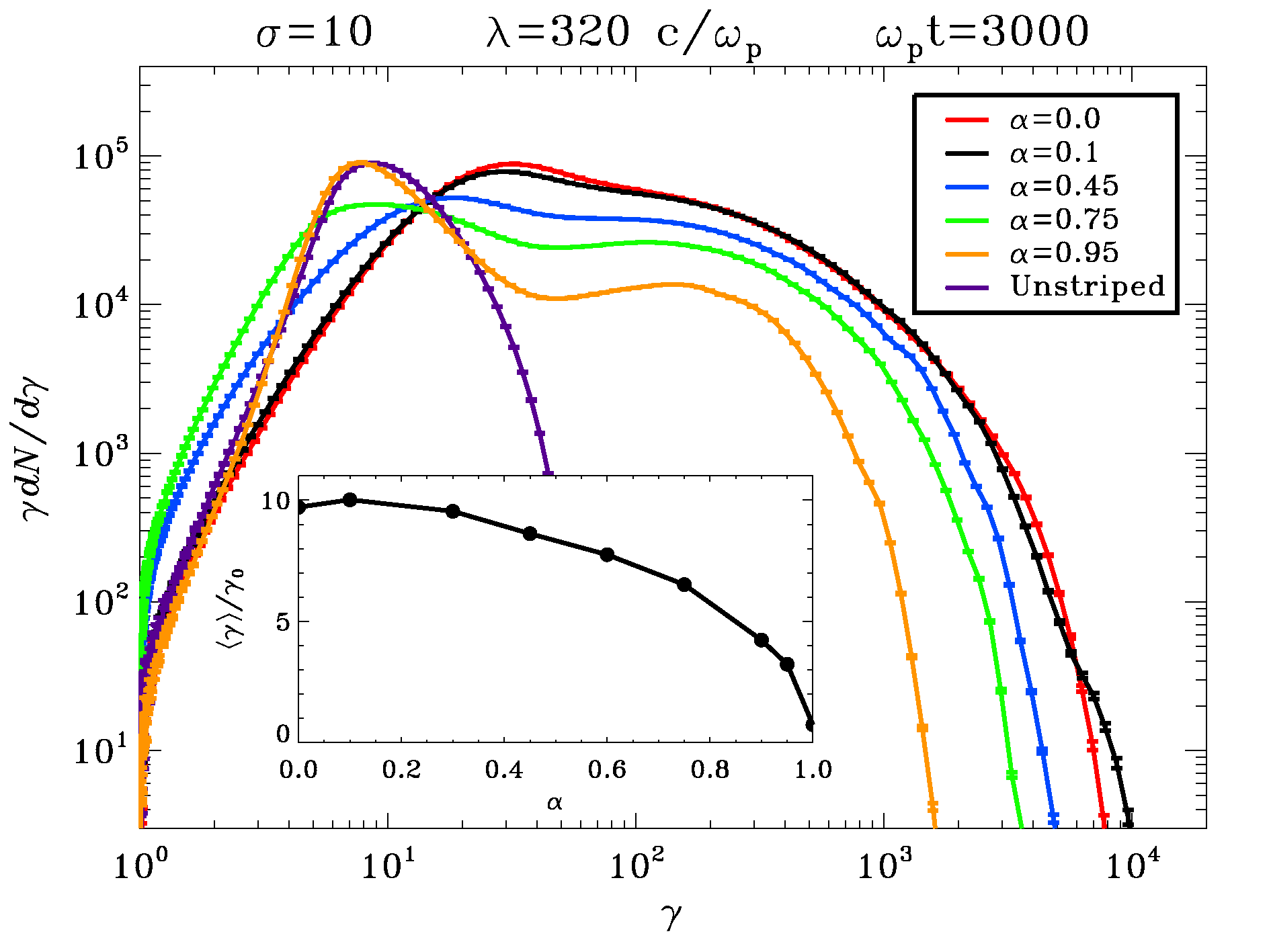}
\caption{Downstream particle spectrum at $\ompt=3000$ for different values of the stripe-averaged field $\langle B_\phi\rangle_\lambda$ (or equivalently, of the parameter $\alpha$), in a flow with $\lambda=320\comp$ and $\sigma=10$, from \citet{sironi_spitkovsky_11b}. The limit of an unstriped wind is shown for reference as a purple line. The black line in the subpanel shows the average downstream Lorentz factor as a function of $\alpha$ (with $\alpha=1.0$ referring to the unstriped wind).}
\label{fig:specdc}
\end{center}
\end{figure}

At higher latitudes, the presence of a non-negligible stripe-averaged field $\langle B_\phi\rangle_\lambda$ inhibits the Fermi process, in analogy to the case of perpendicular magnetized shocks (with uniform fields) discussed at the beginning of this section. The efficiency of particle acceleration via shock-driven reconnection is also affected, as we show in \fig{specdc}. For $\alpha\lesssim0.1$  (i.e., relatively close to the equatorial plane), the field is efficiently dissipated, and the shape of the spectrum is nearly independent of latitude. For $\alpha\gtrsim0.3$, the post-shock particle spectrum consists of two components. The low-energy peak comes from cold plasma with mean Lorentz factor $\sim \gamma_0$, whereas the high-energy part is populated by hot particles that gained energy from field dissipation, so that their mean Lorentz factor is now $\sim \gamma_0 \sigma$. As $\alpha$ increases, the fraction of upstream Poynting flux available for dissipation decreases, which explains why the high-energy component in the spectra of \fig{specdc} gets de-populated, at the expense of the low-energy part. The limit $|\alpha|\rightarrow 1$ (yellow curve for $\alpha=0.95$) approaches the result expected for an unstriped wind (purple line).

Based on these findings, one could interpret the optical and X-ray signatures of the Crab, which require a particle spectrum with $p\simeq2.5$, as synchrotron emission from the particles that are Fermi-accelerated close to the  equatorial plane of the  wind. In addition, the spectral index required for the radio spectrum of the Crab ($p\simeq1.5$) could naturally result from  the  broad hard component of particles accelerated by shock-driven reconnection. However, the particle spectrum in the simulations approaches the hard tail required by the observations only when the combination $\lambda/r_{L,hot}$ exceeds a few tens (for smaller values, the spectrum is a narrow  thermal-like distribution). At the termination shock of the pulsar wind ($R=R_{TS}$) we have
\be\label{eq:wind}
\frac{\lambda}{r_{L,hot}}\simeq4\pi\kappa \frac{R_{LC}}{R_{TS}}~~,
\ee
where $R_{LC}=c/\Omega$ is the light cylinder radius ($\Omega=2\pi/P$ is the pulsar rotational frequency), and $\kappa$ is the so-called multiplicity in the wind (i.e., the ratio of the actual density to the Goldreich-Julian density, \citet{goldreich_julian_69}). For the Crab, $R_{TS}\simeq5\times10^8R_{LC}$ \citep{hester_02} and most available models estimate $\kappa\simeq10^4-10^6$ \citep{bucciantini_11}. Based on our findings,  the resulting value of $\lambda/(r _L \sigma)\lesssim0.01$ would yield a Maxwellian-like spectrum, at odds with the wide flat spectrum required by observations. If radio-emitting electrons are accelerated at the termination shock of pulsar winds via magnetic reconnection, a revision of the existing theories of pulsar magnetospheres is required.

%%%%%%%%%%%%%%%%%%%%%%%%%%%%%%%%%%
\subsection{The flaring emission}\label{sect:flare}
In recent years, the {\it Fermi} and {\it AGILE} satellites have detected a number of hours- to week-long flares at GeV energies, which surprisingly falsify the widely-believed ``standard candle'' nature of the high-energy Crab emission. During these events the Crab nebula gamma-ray flux above 100 MeV exceeded its average value by a factor of several or higher \citep{2011Sci...331..736T, 2011Sci...331..739A,Buehler:2012aa}, while at other wavelengths nothing unusual was observed \citep{Weisskopf:2013aa}. The observed gamma-ray flares  happen with a cadence of $\sim1\,{\rm year}$ \citep[e.g.,][]{Buehler:2014aa} and there are no associated pulsar timing glitches. Variability on timescales as short as a few hours has been reported. The peak isotropic luminosity is roughly $10^{36}\,{\rm ergs/s}$ and the energy radiated is $\sim10^{41}\,{\rm ergs}$. The flares and secular observations \citep{Wilson-Hodge:2011aa} demonstrate that the energy conversion is intermittent and that the mechanism can be locally cataclysmic.

The flare properties suggest that an extreme accelerator is at work. The typical decay time of the flaring episodes, which is attributed to synchrotron cooling, together with the $\sim$ GeV peak frequency, allows to solve simultaneously for the magnetic field strength $\sim 5 $ mG (as compared to the nebula-averaged $\sim 200 \,{\rm \mu G}$) and for the extreme energy of the emitting particles $\sim $ PeV. From the $\sim 10$-hour rise time of the flares, one can estimate the size $\sim 10^{15}$ cm of the emission region. In order to accelerate up to PeV energies within this length, the accelerating electric field needs to be comparable to the inferred magnetic field (i.e., $E\sim B$).

Fermi acceleration at the termination shock of the Crab nebula fails to explain the observed GeV flares \citep{sironi_13}. In contrast, the requirement that $E\sim B$ is naturally satisfied in reconnection layers, in the relativistic regime where the magnetic energy per particle exceeds its rest mass, or equivalently where the magnetization $\sigma=B_0^2/4\pi \rho_0 c^2\gg1$. The reconnection scenario would work best in the most magnetized regions of the nebula, i.e., near the poles and possibly in the jets \citep{2012ApJ...746..148C, 2012MNRAS.427.1497L, 2013MNRAS.428.2459K, 2013MNRAS.436.1102M}. Unfortunately, current gamma-ray telescopes do not have the angular resolution to pin down the precise location of the flares within the Nebula. Below, we discuss how PIC simulations help unveiling the role of magnetic reconnection in the Crab Nebula, as the underlying particle accelerator that powers the GeV flares.

\subsubsection{Plane-parallel reconnection}
In the simplest geometry of magnetic reconnection, the field lines are parallel to a pre-existing current sheet, with opposite polarity on the two sides of the current sheet. We shall call this setup as ``plane-parallel reconnection'' \citep[see][for a review]{Kagan:2015aa}. Using 2D and 3D PIC simulations, it has been recently shown that most of the features of the Crab flares can be explained with relativistic plane-parallel reconnection (timescale, energetics, particle and photon spectra). The key arguments in favor of reconnection for the Crab flares are:

\begin{figure}[tbp]
\begin{center}
\includegraphics[width=0.75\textwidth]{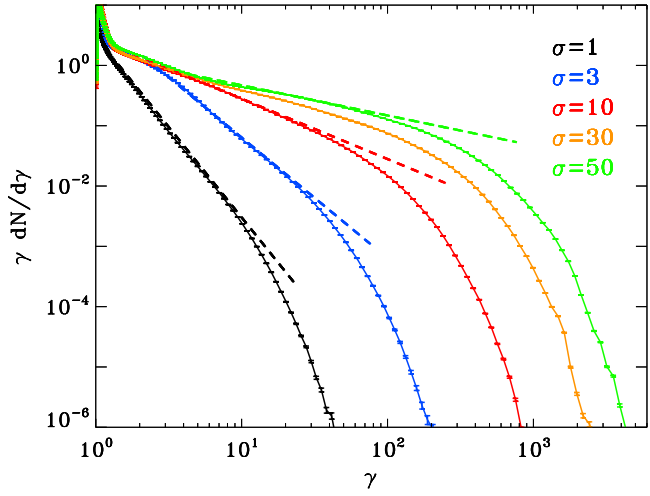}
\caption{Dependence of the spectrum on the magnetization, as indicated in the legend, from \citet{sironi_spitkovsky_14}. The dotted lines refer to power-law slopes of $-4$, $-3$, $-2$ and $-1.5$ (from black to green).}
\label{fig:spec2db}
\end{center}
\end{figure}

\begin{figure}[]
\centering
\includegraphics[width=13cm]{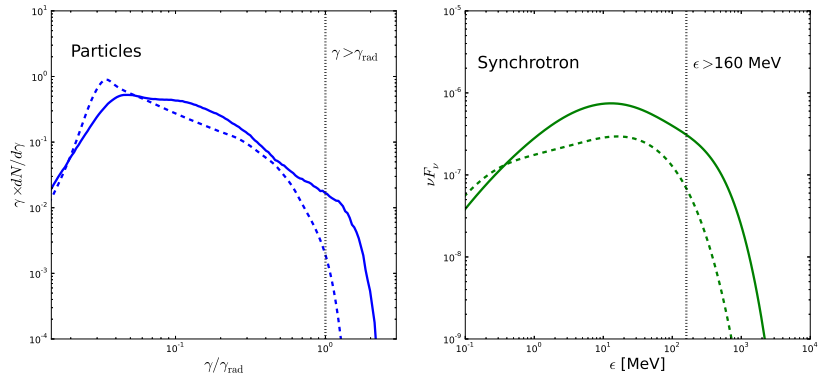}
\caption{Isotropically-averaged particle spectrum ($\gamma d{ N}/d\gamma$, left panel) and synchrotron radiation energy distribution ($\nu F_{\nu}$, right panel) in a 2D (solid line) and 3D (dashed line) PIC simulations of relativistic reconnection, including the effect of the radiation reaction force on the particles. The vertical dotted lines show the radiation-reaction limited energy of a particle if $E=B_0$ ($\gamma=\gamma_{\rm rad}$, left), and the corresponding maximum synchrotron photon energy ($\epsilon=160~$MeV independent of $E$ and $B_0$, right). Figure adapted from \citet{2014PhPl...21e6501C}.}
\label{fig_burnoff}
\end{figure}

\begin{figure}[t]
\begin{center}
\hspace{0cm}
\vspace{0.cm}
\hbox{
\includegraphics[width=1\textwidth]{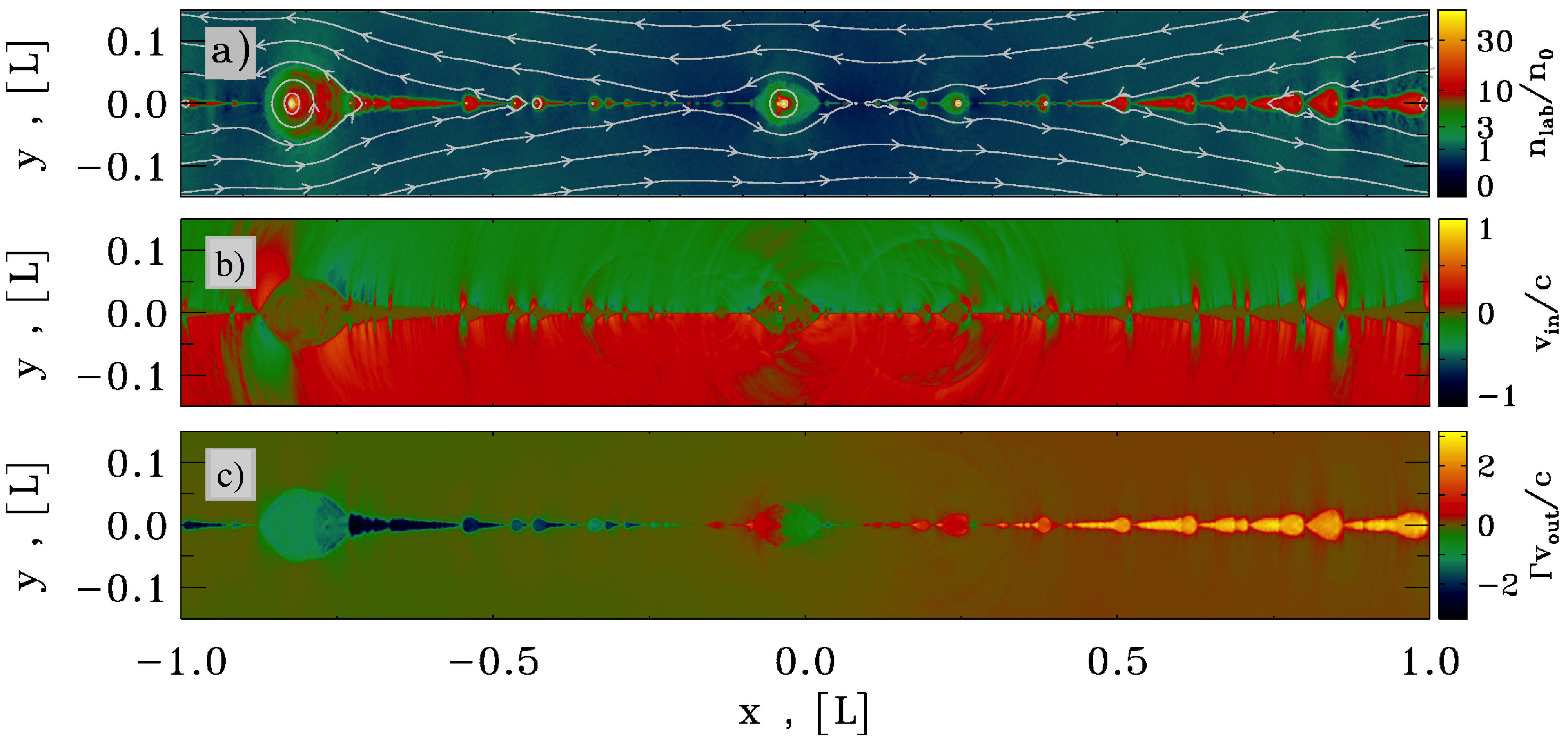}
}
\end{center}
\caption{
\footnotesize {The reconnection layer, from a 2D
    simulation with $\sigma=10$, from \citet{sironi_16}. 
We present (a) particle density, in units of the number
    density far from the sheet, with overplotted magnetic field lines; 
(b)  inflow velocity; and (c)
     outflow four-velocity, in units of the speed of light. The plasma enters the reconnection 
layer with $v_{\rm rec}\sim 0.15\,c$. The reconnection layer fragments into a series of magnetic islands
(or plasmoids), {moving away from the center of the current sheet at ultra-relativistic speeds.}}}
\label{fig:fluid2d}
\end{figure}

\begin{itemize}

\item The flare
spectrum requires an electron power-law population with hard slope $p=-d\log N/d\log\gamma\lesssim 2$, which is not attainable in
shocks, but it naturally results from relativistic reconnection \citep{sironi_spitkovsky_14,Guo:2014aa,melzani14,Werner:2016aa,sironi_15,sironi_16}. As shown in \fig{spec2db}, the power-law slope depends on the flow magnetization, being harder for higher $\sigma$ ($p\sim1.5$ for $\sigma=50$, compare solid and dotted green lines). The slope is steeper for lower magnetizations  ($p\sim4$ for $\sigma=1$, solid and dotted black lines), approaching the result of non-relativistic reconnection, yielding poor acceleration efficiencies \citep[][]{drake_10}.

\item The $\sim \,$GeV peak energy of the flares is well above the classical synchrotron ``burnoff'' limit of $\sim 236 \,\eta$ MeV (as measured in the fluid rest frame), which is obtained by balancing acceleration due to an electric field $E=\eta \,B$ with synchrotron cooling losses; unlike in shocks, where $\eta<1$, in the reconnection layer one finds $\eta>1$, thus boosting the synchrotron limit to the observed $\sim$ GeV peak. In a reconnection scenario, this requires the accelerating particles to stay confined within the reconnection layer, where $\eta>1$. \citet{2011ApJ...737L..40U} showed analytically that as the particle energy increases, the trajectory gets more and more focused along the electric field, with vanishing cooling losses. This has now been confirmed with PIC simulations \citep{Cerutti:2013aa, Cerutti:2014aa}. In particular, these studies demonstrated that reconnection can accelerate particles above the synchrotron radiation burn-off limit \citep{1983MNRAS.205..593G, 1996ApJ...457..253D} deep inside the reconnection layer where the electric field overcome the magnetic field (see Fig.~\ref{fig_burnoff}). This result is crucial because it can explain the emission of $>100~$MeV synchrotron radiation emitted during every Crab flare, which would be impossible to achieve in ideal MHD. 

\item  The short rise time ($\sim 10$ hours) of the flaring episodes can naturally result from the inhomogeneity of the reconnection layer, which is fragmented into a chain of magnetic islands, or plasmoids, as shown in \fig{fluid2d} (in 3D, these plasmoids appear as elongated magnetic flux ropes). The plasmoids are overdense (\fig{fluid2d}a), filled with energetic particles and confined by
strong fields. 
%As a result of the tearing instability, the reconnection
%layer fragments into a series of magnetic islands / plasmoids,
%separated by X-points. Over time, the islands coalesce and grow to
%larger scales.
The plasma flows into the reconnection layer at 
$v_{\rm rec}\simeq 0.15\,c$ for $\sigma=10$ (\fig{fluid2d}b). The inflow speed
is nearly independent of $\sigma$ for larger magnetizations \citep{sironi_16}, in agreement
with analytical models \citep{2005MNRAS.358..113L}.
After entering the sheet, the flow is advected out by the tension force of the reconnected
 field. The motion in the reconnection exhausts is
ultra-relativistic (\fig{fluid2d}c), approaching a bulk four-velocity $\Gamma v_{\rm
  out}\sim \sqrt{\sigma}\,c$, in agreement with
the theory \citep{2005MNRAS.358..113L}. 
The relativistic bulk motion of the plasmoids in the
reconnection layer plays a critical role in enhancing ---  via Doppler boosting --- their emission signatures. 

Aside from bulk Doppler beaming, an energy-dependent ``kinetic'' beaming has also been proposed to explain the extreme time variability of the Crab flares \citep{Cerutti:2013aa, Cerutti:2014aa}. In particular, while low-energy particles are nearly isotropic, at high energies ($\gamma\gtrsim\sigma$) the particles exhibit clear sign of anisotropy with two beams pointing roughly towards the $\pm x$-directions, i.e., along the reconnection exhausts. Hence, the beams are not necessarily pointing along the direction $z$ of the reconnection electric field because the tension of the reconnected field lines pushes the particles away from the X-points in the form of a reconnection outflow towards the magnetic islands. Nonetheless, the direction of the beam of energetic particles is not static: it wiggles rapidly within the $xz$-plane, which results in rapid flares of energetic radiation when the beam crosses the line of sight of a distant observer. Since in the Crab the particles emitting $>100$~MeV synchrotron radiation should be accelerated and radiating over a sub-Larmor timescale, one expects that the highest energy radiation should keep the imprint of the particle anisotropy (regardless of the acceleration process), while the low-energy radiation should be more isotropic \citep{Cerutti:2013aa, Cerutti:2014aa}.

\end{itemize}

\subsubsection{Explosive reconnection}
Despite its successes, plane-parallel reconnection does not seem to be fast enough to explain the short rise time of the Crab flares. In particular, recent PIC studies of relativistic reconnection have demonstrated that the reconnection rate (inflow velocity) in 3D simulations of plane-parallel reconnection is significantly lower than in 2D. For a reference magnetization $\sigma=10$ the reconnection rate in 2D is $v_{rec}/c\sim0.1$, whereas in 3D it is only $v_{rec}/c\sim0.02$ \citep{sironi_spitkovsky_14}. The slower reconnection rate leads to a weaker accelerating electric field. Moreover, for a given flare duration it translates into a smaller utilised magnetic energy. 

To overcome this difficulty, it has recently been proposed that the Crab flares might result from explosive reconnection episodes (a process that has  been called ``magnetoluminescence'' by \citet{blandford_17},  for the rapid conversion of magnetic energy into high-energy particles and then into radiation), accompanying the relaxation of force-free equilibria  on dynamical timescales (i.e., corresponding to an effective reconnection rate of $v_{rec}/c\sim1$). In particular, \citet{Nalewajko:2016aa,Lyutikov:2016aa,Yuan:2016aa} have carried out PIC simulations of the relaxation of force-free equilibria in application to the Crab flares.

As a representative case, we consider the configuration of two
Lundquist's force-free cylinders surrounded by uniform magnetic field \citep{Lyutikov:2016aa},
\be
{\bf B}_L(r\le r_{\rm j}) \propto J_1 (r \alpha_0) {\bf e}_\phi +  J_0 (r \alpha_0) {\bf e}_z\,,
\label{eq:lundquist}
\ee
Here, $J_{0},\,J_{1}$ are Bessel functions of zeroth and first order and the constant $\alpha_0\simeq 3.8317$ is the first root of $J_{0}$.  
This solution is terminated at the first zero of $J_1$, which we denote 
as $r_j$ and hence continued with $B_z=B_z(r_j)$ and $B_\phi=0$ for $r>r_j$. Since the total current of the flux tube is zero, the azimuthal field vanishes at the boundary of the rope, and so the evolution is initially very slow (i.e., the initial configuration is dynamically stable).
 To speed things up, the ropes are pushed towards each other.
 
 \begin{figure}[!]
\centering
\includegraphics[width=.49\textwidth]{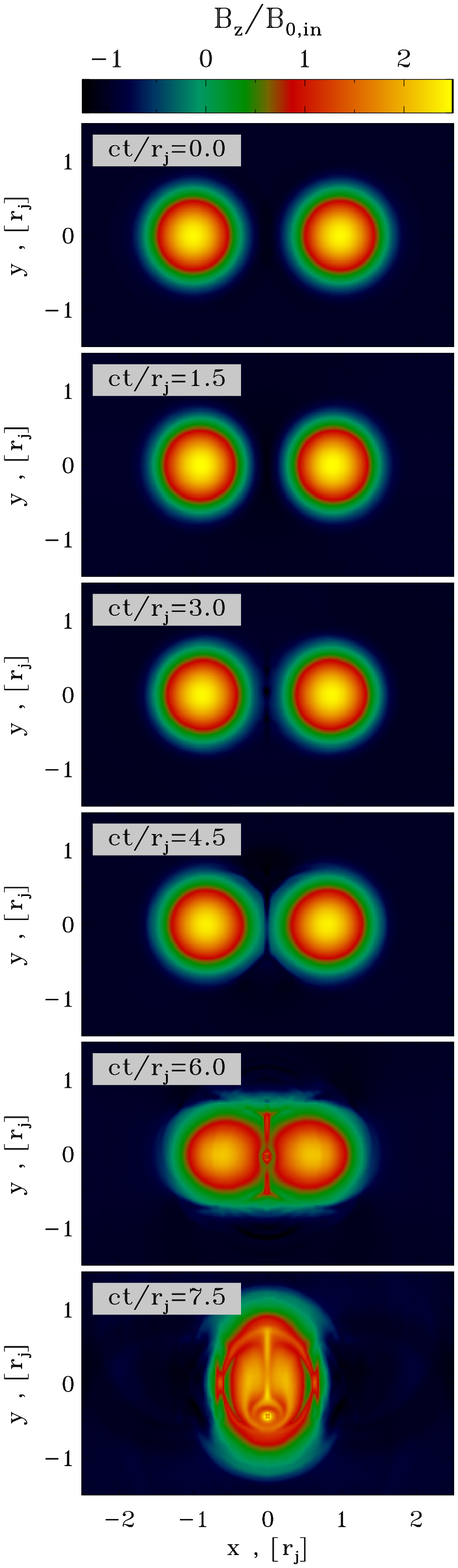} 
\includegraphics[width=.49\textwidth]{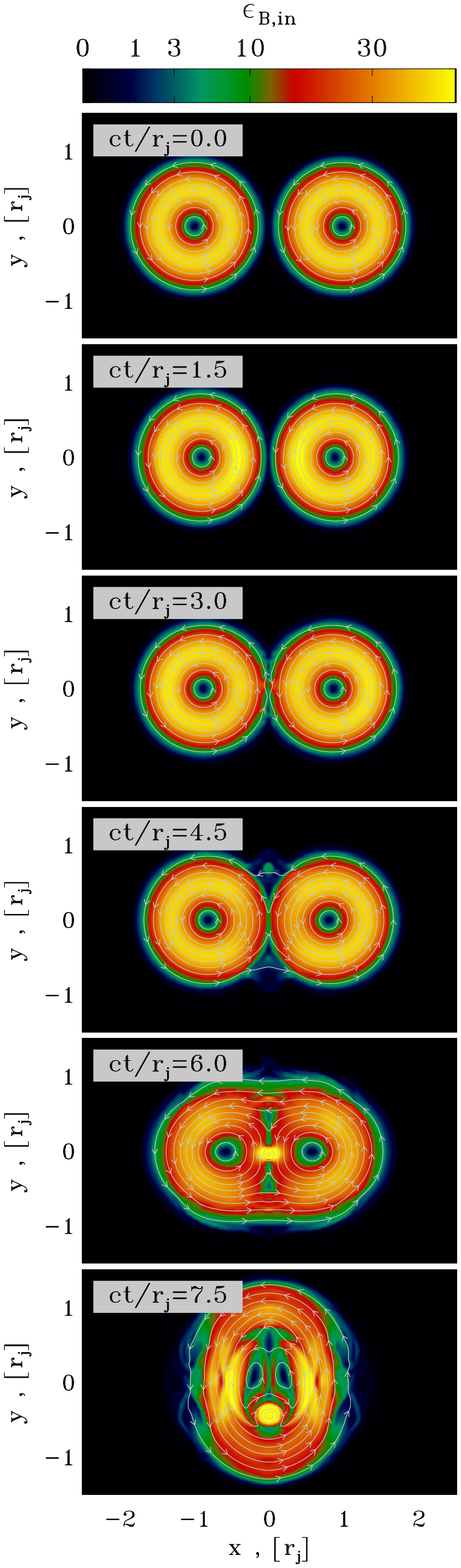} 
\caption{Temporal evolution of 2D Lundquist ropes (time is measured in $c/\rj$ and indicated in the grey box of each panel, increasing from top to bottom), from  \citet{Lyutikov:2016aa}. The plot presents the 2D pattern of the out-of-plane field $B_z$ (left column) and of the in-plane magnetic energy fraction $\epsilon_{B,\rm in}=(B_x^2+B_y^2)/8 \pi n m c^2$ (right column; with superimposed magnetic field lines), from a PIC simulation with $\sigmain=43$ and $\rj=61\,\rhot$.}
\label{fig:lundfluid} 
\end{figure}
\begin{figure}[h!]
\centering
\includegraphics[width=.6\textwidth]{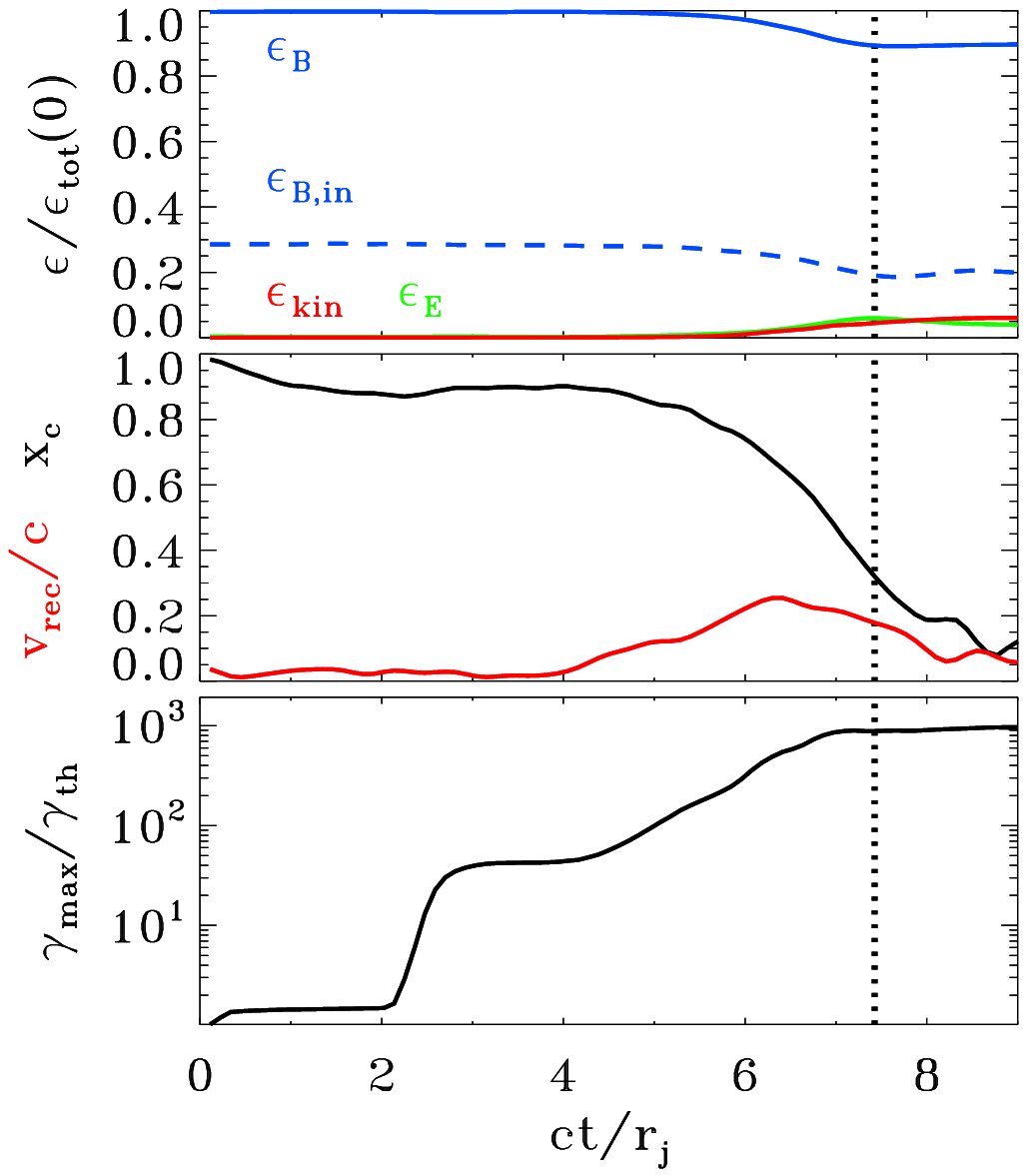} 
\caption{Temporal evolution of various quantities, from a 2D PIC simulation of Lundquist ropes with $\sigmain=43$ and $\rj=61\,\rhot$ (the same as in \fig{lundfluid}), from  \citet{Lyutikov:2016aa}. Top panel: fraction of energy in magnetic fields (solid blue), in-plane magnetic fields (dashed blue), electric fields (green) and particles (red; excluding the rest mass energy), in units of the total initial energy. Middle panel: reconnection rate $v_{\rm rec}/c$ (red), and location $x_{\rm c}$ of the core of the rightmost flux rope (black), in units of $\rj$.
Bottom panel: evolution of the maximum Lorentz factor $\gammamax$.}
\label{fig:lundtime} 
\end{figure}
 
In \fig{lundfluid}, we present the 2D pattern of the out-of-plane field $B_z$ (left column) and of the in-plane magnetic energy fraction $\epsilon_{B,\rm in}=(B_x^2+B_y^2)/8 \pi n m c^2$ (right column; with superimposed magnetic field lines), from a PIC simulation with  $\sigmain=43$ (only defined with the in-plane fields) and $\rj=61\,\rhot$ (where $\rhot$ is the Larmor radius of particles heated by reconnection).
As the two magnetic ropes slowly approach, driven by the initial velocity push, reconnection is triggered in the plane $x=0$, as indicated by the formation and subsequent ejection of small-scale plasmoids. 
%Until $ct/\rj\sim 4.5$, the cores of the two islands have not significantly moved (black line in the middle panel of \fig{lundtime}, indicating the $x_{\rm c}$ location of the center of the rightmost island), the reconnection speed is quite small (red line in the middle panel of \fig{lundtime}) and no significant energy exchange has occurred from the fields to the particles (compare the in-plane magnetic energy, shown by  the dashed blue line in the top panel of \fig{lundtime}, with the particle kinetic energy, indicated with the red line). 
As a result of reconnection, an increasing number of field lines, that initially closed around one of the ropes, are now engulfing both magnetic islands. Their tension force causes the two ropes to approach and merge on a quick (dynamical) timescale, starting at $ct/\rj\sim 4.5$ and ending at $ct/\rj\sim 7.5$ (see that the distance of the rightmost island from the center rapidly decreases, as indicated by the black line in the middle panel of \fig{lundtime}). The tension force drives the particles in the flux ropes toward the center, with a fast reconnection speed peaking at $v_{\rm rec}/c\sim 0.3$ (red line in the middle panel of \fig{lundtime}).\footnote{The reconnection rate is measured to be in the range $v_{\rm rec}/c\sim0.2-0.5$, which increases with the magnetization and saturates at around 0.5 at high magnetization limit \citep{Lyutikov:2016aa}.} The reconnection layer at $x=0$  stretches up to a length of $\sim 2\rj$, and secondary plasmoids are formed. 
%In the central current sheet, it is primarily the in-plane field that gets dissipated (compare the dashed and solid blue lines in the top panel of \fig{lundtime}), driving an increase in the electric energy (green) and in the particle kinetic energy (red). 
In this phase of evolution, the fraction of initial energy released to the particles is small ($\epsilon_{\rm kin}/\epsilon_{\rm tot}(0)\sim 0.1$, top panel in \fig{lundtime}), but the particles advected into the central X-point experience a dramatic episode of acceleration. As shown in the bottom panel of \fig{lundtime}, the cutoff Lorentz factor $\gammamax$ of the particle spectrum presents a dramatic evolution, increasing up to $\gammamax/\gamma_{\rm th}\sim 10^3$ within a couple of dynamical times (here, $\gamma_{\rm th}$ is the initial ``thermal'' Lorentz factor). 

\begin{figure}[h!]
\centering
\includegraphics[width=.75\textwidth]{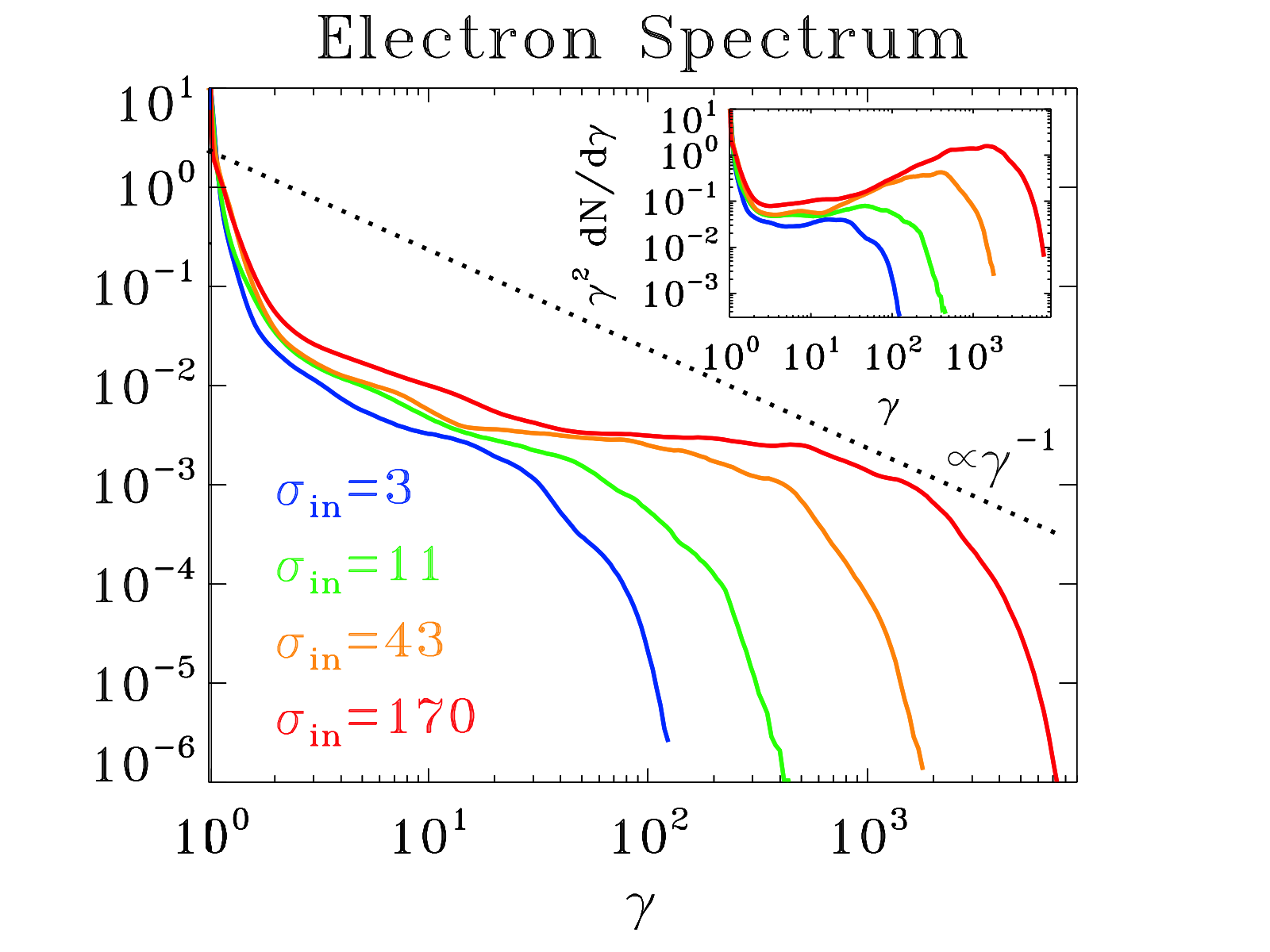} 
\caption{Particle spectrum for a suite of PIC simulations of Lundquist ropes, from \citet{Lyutikov:2016aa}. We fix $\rj/\rhot=61$ and we vary the magnetization $\sigmain$ from 3 to 170 (from blue to red, as indicated by the legend). The main plot shows $\gamma dN/d\gamma$ to emphasize the particle content, whereas the inset presents $\gamma^2 dN/d\gamma$ to highlight the energy census. The dotted black line is a power law $\gamma dN/d\gamma\propto \gamma^{-1}$, corresponding to equal energy content per decade (which would result in a flat distribution in the insets). The spectral hardness is strongly dependent on $\sigmain$, with higher magnetizations giving harder spectra.}
\label{fig:lundspeccomp} 
\end{figure}

This phase of extremely fast particle acceleration  on a dynamical timescale is analogous to the relaxation of unstable ``ABC'' force-free structures discussed by \citet{Nalewajko:2016aa,Lyutikov:2016aa}, and it constitutes the most promising scenario to explain the Crab flares.  
The particle acceleration efficiency  and the hardness of the power-law slope depend on the mean magnetization of the configuration, in a similar fashion as in plane-parallel reconnection scenarios. The particle spectrum gets harder as the mean magnetization increases (\fig{lundspeccomp}); both the non-thermal particle fraction and the maximum particle energy increase with the magnetization \citep{Lyutikov:2016aa,Nalewajko:2016aa}.

Since the highest energy particles are first accelerated in the current layers by the parallel electric field, they do not radiate much when they are inside the sheet, because the curvature of their trajectory is small. Most of the radiation is produced when particles are ejected from the current layers --- their trajectories start to bend significantly in the ambient magnetic field  \citep{Yuan:2016aa}. Such a separation of acceleration site and radiative loss site facilitates acceleration beyond the synchrotron radiation reaction limit, as required by the Crab flares. 
Fast variability of the observed photon flux can be produced when compact plasmoids that contain high-energy particles are ejected from the ends of the current layers and get destroyed. These give beamed radiation. An observer sees high intensity radiation when the beam happens to be aligned with the line of sight. Such peaks in emission are accompanied by an increase in the polarization degree and rapid change of polarization angle in the high-energy band \citep{Yuan:2016aa}. The variability timescale is determined by the spatial extent of the emitting structure, e.g. the plasmoids, thus can be much shorter than the light crossing time of the region that collapses.

%%%%%%%%%%%%%%%%%%%%%%%%%%%
\section{Conclusions}\label{sect_conc}
In this chapter, we have discussed the role of PIC simulations in unveiling the origin of the emitting particles in PWNe. After describing the basics of the PIC technique, we have summarized its implications for the quiescent and flaring emission of the Crab Nebula, as a prototype of PWNe. A consensus seems to be emerging that, in addition to the standard scenario of particle acceleration via the Fermi process at the termination shock of the pulsar wind, magnetic reconnection in the wind, at the termination shock and in the nebula plays a major role in powering the multi-wavelength emission signatures of PWNe.

\begin{acknowledgement}
LS acknowledges support from DoE DE-SC0016542 and NASA Fermi NNX16AR75G. BC acknowledges support from CNES and Labex OSUG@2020 (ANR10 LABX56).
\end{acknowledgement}

\bibliographystyle{aps-nameyear}          % American Physical Society (APS)
\bibliography{pic_pwn}

\end{document}